\documentclass[12pt,epsf]{article}
\usepackage{latexsym}

\newcommand{\rf}[1]{(\ref{#1})}
\newcommand{\bea}{\begin{eqnarray}}
\newcommand{\eea}{\end{eqnarray}}

\renewcommand{\d}{\mbox{d}}
\newcommand{\g}{\gamma}

\renewcommand{\a}{\alpha}

\newcommand{\om}{\omega}
\newcommand{\del}{\delta}

\newcommand{\sg}{\sigma}
\newcommand{\ra}{\rangle}
\newcommand{\la}{\langle}

\def\void{}
\def\labelmark{}

\newenvironment{formula}[1]{\def\labelname{#1}
\ifx\void\labelname\def\junk{\begin{displaymath}}
\else\def\junk{\begin{equation}\label{\labelname}}\fi\junk}%
{\ifx\void\labelname\def\junk{\end{displaymath}}
\else\def\junk{\end{equation}}\fi\junk\labelmark\def\labelname{}}

{\ifx\void\labelname\def\junk{\end{array}\end{displaymath}}
\else\def\junk{\end{array}\right.\end{equation}}
\fi\junk\labelmark\def\labelname{}\def\junk{}
}

\newcommand{\beq}{\begin{formula}}
\newcommand{\eeq}{\end{formula}}
\newcommand{\beqv}{\begin{formula}{}}

\begin{document}

%\hfill    NBI-HE-2000-xxx

\hfill January 2001

\begin{center}
\vspace{24pt}
{\large \bf High-temperature, classical, real-time dynamics of non-abelian 
gauge theories as seen by a computer}

\vspace{36pt}

{\sl J. Ambj\o rn}$^{a}$, {\sl K.N. Anagnostopoulos}$^{b}$ and
{\sl A. Krasnitz}$^{c}$

\vspace{24pt}\noindent
$^a$ The Niels Bohr Institute, University of Copenhagen,\\
Blegdamsvej 17, DK-2100 Copenhagen \O, Denmark.\\
{E-mail: ambjorn@nbi.dk} \\

\vspace{12pt}\noindent
$^b$ Department of Physics, University of Crete,\\
{P.O. Box 2208, GR-71003 Heraklion, Greece}.\\
E-mail: konstant@physics.uoc.gr

\vspace{12pt}\noindent
$^c$ CENTRA and Faculdade de Ci\^encias e Tecnologia,\\
Universidade do Algarve, \\
Campus de Gambelas, P--8000, Faro, Portugal. \\
{E-mail: krasnitz@ualg.pt}  
  
\end{center}
\vspace{24pt}

\vfill

\begin{center}
{\bf Abstract}
\end{center}
\vspace{12pt}

\noindent
We test at the electroweak scale the recently proposed
elaborate theoretical scenario 
for real-time dynamics of non-abelian gauge theories
at high temperature.
We see no sign of the predicted behavior. This indicates 
that perturbative concepts like color conductivity and
Landau damping might be irrelevant at temperatures corresponding 
to the electroweak scale.

\vspace{12pt}

\bigskip



\vfill

\newpage

\section{Introduction}

The study of $SU(2)$ gauge theories at high temperatures is important 
for our understanding of the electroweak theory in the early universe.
One effect which has been of interest in this context is the 
baryon-number non-conservation in the electroweak theory,
caused by the anomaly of the baryonic current. After the work
of Kuzmin, Rubakov and Shaposhnikov \cite{krs} it was realized that 
the baryon number non-conservation,
caused by thermal field fluctuations between gauge-equivalent
vacua with different winding numbers, could be large in the 
unbroken phase of the electroweak theory.
A quantitative verification
of the KRS-scenario requires non-perturbative real time simulations 
of hot thermal gauge theories, a task we still do not know how 
to perform from first principles. However, following suggestions
by \cite{grs} that one could use {\it classical} thermal gauge theory 
to address the question, it was shown in \cite{aaps} that 
transitions between vacua with different winding numbers
is unsuppressed at high temperatures in the unbroken phase 
of the electroweak theory. Starting with \cite{su2rate,thalgs}, 
much work has since
gone into refining the numerical techniques used and 
turning the qualitative statements into 
quantitative measurements 
\cite{quant1,quant2,quant3,quant4,quant5,quant6,quant7}
and also in understanding 
to what extent the notion of topology used in the continuum
anomaly calculations is still valid in the lattice simulations \cite{topology}.

In order to address the determination of the rate of 
``sphaleron'' transitions quantitatively one needs either 
to understand the corrections to the rate induced 
by using classical thermodynamics rather than the 
correct quantum field theory thermodynamics, or to 
develop better non-perturbative methods suited 
for real time simulations. The latter alternative is of course
preferable since it might allow us to address many other 
non-perturbative questions which involve real-time 
dynamics of non-abelian gauge theories at high temperatures, by
correctly incorporating thermal fluctuations in the ultraviolet.

Much progress has been made in this direction over the 
last decade, starting with the concept of hard thermal loops, and 
culminating with the effective small-momentum, low-frequency 
theory of B\"{o}deker \cite{bodeker} and its interpretation in terms 
of color conductivity \cite{asy,as}. In ordinary (abelian)  plasma physics
low frequency magnetic fields decay slower than naively expected
due to Landau damping. It is now understood that the same mechanism 
applies to a pure non-abelian plasma, where the hard, high frequency modes
couple to the low frequency magnetic modes much like the charged particles
in an abelian plasma, and in this way produce a new time scale 
for the magnetic fluctuations. While the dominant long-range 
magnetic fluctuations will occur at the (non-perturbative) length scale
of the order $ 1/g^2T$, the lifetime of the fluctuations 
will be of the order $ \om \sim g^4T \ln(1/g)$.

On a more quantitative level the soft classical 
fields (momentum $k \leq gT$) couples to hard currents 
according to 
\beq{1.1}
\dot{\mathbf{E}}= {\mathbf D}\times{\mathbf B}- {\mathbf J}_{\rm hard},
\eeq
where 
\beq{1.2}
{\mathbf J}_{\rm hard}= \sg {\mathbf E} + \mbox{\boldmath $\xi$},
\eeq
and where the effective noise term {\boldmath $\xi$} is determined
by the fluctuation-dissipation theorem:
\beq{1.3}
\la \xi^a_i (t,{\mathbf x}) \xi^b_j(t',{\mathbf x'})\ra 
= 2 T \sg \,\del_{ij}\del^{ab} \del(t-t')\del({\mathbf x}-{\mathbf x'}).
\eeq
In \rf{1.1} $\sg$ denotes the so-called color conductivity,
which to a leading $\log$ approximation is given by 
\beq{1.4}
\sg = \frac{m^2}{\g},~~~~m^2 = \frac{2}{3} \,(gT)^2,~~~\g = \frac{3}{16\pi}
\,g^2 T \, \ln (1/g).
\eeq
$m$ denotes the Debye screening mass and $\g$ the hard gauge fields damping
rate (given here for pure $SU(2)$ gauge theory). 
The derivation of \rf{1.1} is valid in a frequency range
\beq{1.5}
\om \ll k \\ \g~~~~{\rm and}~~~~k \geq g^2T .
\eeq
Strictly speaking, in order for \rf{1.5} to be valid, one has formally
to require that $\ln 1/g$ be parametrically large.
In this frequency range one can ignore the time derivative in \rf{1.1}. Then
\beq{1.5a}
{\mathbf D}\times{\mathbf B}= \sg{\mathbf E} + \mbox{\boldmath $\xi$}.
\eeq

Working in temporal gauge  {\it and ignoring non-linearities}, one obtains from
\rf{1.5a}
\beq{1.6}
k^2 {\mathbf A} + \sg \dot{\mathbf{A}} = \mbox{\boldmath $\xi$},
\eeq
which leads to the decay of gauge correlators:
\beq{1.7}
\la{\mathbf A}(t){\mathbf A}(0)\ra  \propto \exp\Big\{-\frac{k^2}{\sg} t\Big\}.
\eeq
The decay time $\tau$ for $k \sim g^2T$ is thus of the order 
\beq{1.7a}
\tau \sim \frac{1}{g^4 T \ln (1/g)},
\eeq
{\it i.e.,} much longer than the non-perturbative length scale $1/(g^2 T)$.

While this picture no doubt gives an appropriate description of the 
long wavelength, low frequency 
thermal fluctuations for spatial scales up to $1/(g^2T)$ and 
time scales up to $1/(g^4T \ln (1/g))$ when $\ln (1/g)$ in some 
sense can be considered large,
the question arises how well it describes ``sphaleron'' physics at the 
electroweak scale where $\a \approx 1/30$, i.e.\ $g \approx 0.65$.

Presently there are good indications that the picture works  
well at the above value of $g$. Real time computer simulations
have been done, using various ways of implementing the hard currents
\cite{quant1,quant3,quant5}
The primary observable in the simulations has been the sphaleron rate.
This is so for good reasons. Firstly, it is an important observable which 
might play a role in explaining the  
baryon asymmetry of the universe. Secondly, it is believed to 
be dominated by long wavelength, classical thermal 
field fluctuations, {\it i.e.},\ one can hope that the classical theory,
or at least the improvements, like B\"{o}deker's effective field theory,
will allow us to determine this (non-perturbative) rate by computer 
simulations. Until now all computer simulations basically agree, and 
there is a kind of consensus in the community that the 
sphaleron rate goes as predicted by Arnold, Son and Yaffe 
(ASY) and B\"{o}deker.

The purpose of this article is not to cast doubt on
the effective theory of B\"{o}deker and ASY, but only 
to point out that it is may not be clear that it can be applied 
to the electroweak theory for temperatures around the 
phase transition. The derivation of the effective theory,
in particular in the framework of ASY, uses perturbative concepts
like real time gauge correlators and color conductivity, 
and the explicit behavior of these objects as functions 
of momemtum $k$ and frequency $\om$ are used. While some of these 
objects are gauge dependent they can nevertheless be measured 
in the same computer simulations used to determine the sphaleron
rate and one can directly check if their long wavelength, low frequency
part show the behavior predicted by the effective theory. If not, 
one could be tempted to conclude that one needs to go to higher 
temperature, and that one should {\em not} try to match the 
sphaleron rate to formulas  based on the validity of the 
long wavelength, low frequency effective theory.

The rest of this paper is organized as follows: in Section 2 we 
describe shortly our simulation setup and the set of 
gauge-covariant and gauge-invariant objects likely
to be governed by a long wavelength effective theory. 
Our results are reported in Section 3.
Section 4 concludes.

\section{Simulation setup and observables}

As mentioned in the Introduction, there presently exists no
genuine non-perturbative way of simulating real-time processes 
in non-abelian  gauge theories from first principles.
Solving the classical field equations of motion
is as close as one can get to to a study from first principles, in the sense 
that no external parameters are introduced into the system. The caveat 
is, of course, that the classical thermal distribution is incorrect: while 
the thermal fluctuations in the quantum theory are cut off at 
scale $T$, the thermal fluctuation of the classical theory is 
only cut off at the cut-off scale $\Lambda \sim 1/a$, where $a$ 
is the lattice spacing. The non-perturbative, long-distance
magnetic sector of the 
thermal theory is not expected to be very sensitive to an exact location
of the ultraviolet cutoff, as long as the cutoff is not too close
to the magnetic scale. At the same time,
the perturbative sector will be changed and, whenever one 
encounters the Debye mass one has to make a replacement:
\beq{2.1}
m^2_D \sim  g^2 T^2 \to g^2 T /a.
\eeq

A typical high-temperature field configuration 
will be dominated by its large-momentum
components, {\it i.e.},\ for the classical field theory
the components of the order of the cut off, for the 
full quantum theory the components of order $T$.
If one is interested in the dynamic behavior
of the system at large distances (from the magnetic screening length and above)
one needs a way to filter out the prevalent large-momentum components of the
field.
One way to do so 
is by cooling. In the context of sphaleron transitions
the method was first applied in the very first numerical work 
on the sphaleron transitions at the electroweak phase transition,
\cite{aals}, as well as in the study of
the flow of eigenvalues of the Dirac operator in 
presence of lattice sphalerons \cite{topology}. Later it was introduced
as a tool to reduce the large momentum, short distance lattice 
artifacts of the sphalerons in  real time simulations 
\cite{quant4}.
As discussed in \cite{quant4}, moderate cooling is best suited for this purpose,
since it leads to an exponential decay of high-frequency modes. In this 
way the sphaleron profile, buried in the thermal fluctuations, will 
be enhanced.  However, one can go a step further and simply view 
the cooling as a general procedure for integrating out the large 
momentum components of the thermal classical field theory, basically 
leaving us with the large distance physics of the classical  theory.

Our procedure is thus the following: first we generate a thermal
field configuration for an $SU(2)$ classical Yang-Mills theory on a lattice
\cite{thalgs}, for a given temperature $T$.
We then let the system evolve according to the classical equations of motion.
At selected instances along the classical trajectory we extract long-wavelength
information from the field configuration by cooling (we defer to the next 
section the discussion of how deep the cooling should be).
Using the new configurations
we can measure whatever observables we have in mind, and since we are 
probing the low momentum, classical sector of the theory, we should be 
able to compare our results with the prediction from 
effective theory given by \rf{1.1}-\rf{1.2} provided we 
make the appropriate substitutions like \rf{2.1} and provided that 
we are in a coupling constant and temperature region where 
\rf{1.1} and \rf{1.2} are valid. 
  
The classical
Hamiltonian dynamics is most conveniently studied in the temporal gauge where 
the electric field ${\mathbf E}({\mathbf x},t)$ is the conjugate momentum to 
${\mathbf A}({\mathbf x},t)$.
It is worth pointing out that this is also the gauge where eqs.\ 
\rf{1.1}-\rf{1.3} naively make sense. As explained in \cite{as,arnold1}
eqs.\ \rf{1.1}-\rf{1.3} can be generalized to other gauges, but we will
not need it here. We will  study unequal-time correlators of the form
\beq{2.a1}
C_{12}(t) \equiv \frac{1}{V} \int \d^3 x \;
\la {\cal O}_1({\mathbf x},t) {\cal O}_2({\mathbf x},0)\ra,
\eeq
where 
${\cal O}_{1,2}$ are designed to probe the long-wavelength properties of 
interest. In particular, if one is interested in the magnetic degrees of
freedom, a natural choice would be the autocorrelators of the magnetic field
tensor and of its covariant curl:
\beq{2.a2}
{\cal O}_{1}({\mathbf x},t) = {\cal O}_{2}({\mathbf x},t)
=\left\{ \matrix{{\mathbf D}\times {\mathbf B}({\mathbf x},t) \cr
{\mathbf B}({\mathbf x},t)}
   \right.
\eeq
There is a special choice of ${\cal O}_{1,2}$ that allows to determine
the color conductivity $\sigma$. Consider a correlator of
${\bf B}({\bf x},0)$ with \rf{1.5a} at time $t$.  It follows that
\beq{2.a3}
\sg(t) \equiv\frac{\int \d^3 x\;\langle {\bf D}\times{\bf B}({\bf x},0)\cdot{\bf D}\times{\bf B}({\bf x},t)\rangle}{\int \d^3 x \;\langle {\bf D}\times{\bf B}({\bf x},0)\cdot {\bf E}({\bf x},t)\rangle}\longrightarrow\sigma,
\eeq
if the time lag $t$ is large compared to the autocorrelation time of the 
effective noise. 

On the other hand, we can also study the autocorrelator of the color charge 
density
\beq{2.a4}
\rho \equiv     \mathbf{D\cdot E},
\eeq
if we are interested in detecting soft longitudinal excitations of the system
(plasmons).

Note that all the space-local correlators of gauge-covariant quantities are 
invariant
under the residual time independent
gauge transformations which are not fixed by the choice of temporal 
gauge.
However, if we transform these correlators away from the temporal gauge,
they will in general depend on a Wilson line in the time direction, connecting
${\bf x},0$ to ${\bf x},t$. This being the case, two remarks are in order.
First,
it is unclear to us to what extent \rf{2.a3} can serve as 
a genuine gauge invariant definition of color conductivity,
since it is based on correlations between gauge covariant
objects.  
In fact, as emphasized in \cite{as} it is not 
very clear even in a perturbative framework how to define color
conductivity beyond leading perturbative order. In \cite{as} it was 
defined to next to leading order simply as the coefficient which 
appears in the effective theory \rf{1.1}-\rf{1.3}, when formulated 
in temporal gauge or related, so called flow gauges, of which the
Coulomb gauge is a limiting case. Our definition can be seen as an 
extrapolation of this philosophy: we have defined an effective theory
by integrating out the high momentum part of the (classical) thermal
theory and we then define $\sg$ by \rf{1.1} and \rf{1.2}. Then a 
measurement of $\sg(t)$ by \rf{2.a3} can be viewed as testing 
if our effective theory looks anything like \rf{1.1} and \rf{1.2}.

Secondly, it is not apriori clear what effect the connecting temporal Wilson
line has on the characteristic time scale of correlators between 
gauge-covariant quantities. For this reason, we also study the dynamics of a
truly gauge-invariant object, $B^2({\bf x},x)$. As our results show, there is 
hardly any difference in correlation time scales for gauge-covariant and
for gauge-invariant quantities.

\section{Numerical results}

A real-time correlator $C(t)$ of cooled classical gauge fields on a lattice 
depends on the following dimensional parameters: the temperature and the 
coupling constant in the unique combination $g^2T$, the system size $L$, the 
cooling time $\tau$, and, finally, the lattice spacing $a$. As usual, we express
all the other dimensional quantities in terms of the lattice spacing. The choice
of other parameters is dictated by physical considerations. In particular, the
inverse lattice temperature $\beta\equiv 4/(g^2Ta)$ is chosen within a range 
where the ratio of the perturbative Debye mass $m_D$ to $g^2T$ of the classical
theory is close to that of the full SU(2) Yang-Mills theory at electroweak 
temperatures $T\sim 100$GeV. Since we are interested in the dynamics of fields
with momenta of the order of $g^2T$, the dimensionless combination $L/(\beta a)$
should be large enough in order to avoid finite-size effects. Most of our 
simulations were performed at $L/(\beta a)=2.4$. We verified that variations
of $L/(\beta a)$ around that value did not have a measurable effect. Finally, 
the cooling time $\tau$ should be large enough in order to suppress modes with
momenta harder than those on the $g^2T$ scale. In most our simulations 
$(g^2T)^2\tau=3.84$. We saw that further increase of $\tau$ had little impact on
the real-time behavior of the correlators. 

Our original motivation for measuring real-time correlators of cooled fields was
to determine the color conductivity, as explained in the previous section. We
will discuss our (thus far unsuccessful) attempt to determine $\sigma$ at the
end of this section. However, the real-time correlators we measure are 
interesting in their own right, and we will first describe their properties
as they transpired in the simulation. 

In Figures \ref{bbrr} through \ref{plcrr} we present the autocorrelators 
$\langle {\bf D}\times{\bf B}({\bf x},t)\cdot{\bf D}\times{\bf B}({\bf x},0)
\rangle$, $\langle {\bf B}({\bf x},t)\cdot{\bf B}({\bf x},0)\rangle$, and
$\langle B_i^2({\bf x},t)B_i^2({\bf x},0)\rangle-
\langle B_i^2({\bf x},t)\rangle\langle B_i^2({\bf x},0)\rangle$, respectively.
Note that in
the first two cases we determine autocorrelators of gauge-covariant quantities,
whereas in the third case the quantity in question is gauge-invariant.  In each 
case the autocorrelators are normalized by their value at the origin, while the 
time variable is expressed in units of $4/(g^2T)$. Notably, in all three cases
the curves corresponding to different values of $\beta$ coincide as long as the
correlators retain a substantial fraction of their original value. This property
is especially evident in the first two cases, where the error bars are smaller.
\begin{figure}[hbt]
%\input bbrr
% GNUPLOT: LaTeX picture
\setlength{\unitlength}{0.240900pt}
\ifx\plotpoint\undefined\newsavebox{\plotpoint}\fi
\sbox{\plotpoint}{\rule[-0.200pt]{0.400pt}{0.400pt}}%
\begin{picture}(1500,900)(0,0)
\font\gnuplot=cmr10 at 10pt
\gnuplot
\sbox{\plotpoint}{\rule[-0.200pt]{0.400pt}{0.400pt}}%
\put(161.0,123.0){\rule[-0.200pt]{4.818pt}{0.400pt}}
\put(141,123){\makebox(0,0)[r]{0}}
\put(1419.0,123.0){\rule[-0.200pt]{4.818pt}{0.400pt}}
\put(161.0,257.0){\rule[-0.200pt]{4.818pt}{0.400pt}}
\put(141,257){\makebox(0,0)[r]{0.2}}
\put(1419.0,257.0){\rule[-0.200pt]{4.818pt}{0.400pt}}
\put(161.0,391.0){\rule[-0.200pt]{4.818pt}{0.400pt}}
\put(141,391){\makebox(0,0)[r]{0.4}}
\put(1419.0,391.0){\rule[-0.200pt]{4.818pt}{0.400pt}}
\put(161.0,525.0){\rule[-0.200pt]{4.818pt}{0.400pt}}
\put(141,525){\makebox(0,0)[r]{0.6}}
\put(1419.0,525.0){\rule[-0.200pt]{4.818pt}{0.400pt}}
\put(161.0,659.0){\rule[-0.200pt]{4.818pt}{0.400pt}}
\put(141,659){\makebox(0,0)[r]{0.8}}
\put(1419.0,659.0){\rule[-0.200pt]{4.818pt}{0.400pt}}
\put(161.0,793.0){\rule[-0.200pt]{4.818pt}{0.400pt}}
\put(141,793){\makebox(0,0)[r]{1}}
\put(1419.0,793.0){\rule[-0.200pt]{4.818pt}{0.400pt}}
\put(161.0,123.0){\rule[-0.200pt]{0.400pt}{4.818pt}}
\put(161,82){\makebox(0,0){0}}
\put(161.0,840.0){\rule[-0.200pt]{0.400pt}{4.818pt}}
\put(344.0,123.0){\rule[-0.200pt]{0.400pt}{4.818pt}}
\put(344,82){\makebox(0,0){0.4}}
\put(344.0,840.0){\rule[-0.200pt]{0.400pt}{4.818pt}}
\put(526.0,123.0){\rule[-0.200pt]{0.400pt}{4.818pt}}
\put(526,82){\makebox(0,0){0.8}}
\put(526.0,840.0){\rule[-0.200pt]{0.400pt}{4.818pt}}
\put(709.0,123.0){\rule[-0.200pt]{0.400pt}{4.818pt}}
\put(709,82){\makebox(0,0){1.2}}
\put(709.0,840.0){\rule[-0.200pt]{0.400pt}{4.818pt}}
\put(891.0,123.0){\rule[-0.200pt]{0.400pt}{4.818pt}}
\put(891,82){\makebox(0,0){1.6}}
\put(891.0,840.0){\rule[-0.200pt]{0.400pt}{4.818pt}}
\put(1074.0,123.0){\rule[-0.200pt]{0.400pt}{4.818pt}}
\put(1074,82){\makebox(0,0){2}}
\put(1074.0,840.0){\rule[-0.200pt]{0.400pt}{4.818pt}}
\put(1256.0,123.0){\rule[-0.200pt]{0.400pt}{4.818pt}}
\put(1256,82){\makebox(0,0){2.4}}
\put(1256.0,840.0){\rule[-0.200pt]{0.400pt}{4.818pt}}
\put(1439.0,123.0){\rule[-0.200pt]{0.400pt}{4.818pt}}
\put(1439,82){\makebox(0,0){2.8}}
\put(1439.0,840.0){\rule[-0.200pt]{0.400pt}{4.818pt}}
\put(161.0,123.0){\rule[-0.200pt]{307.870pt}{0.400pt}}
\put(1439.0,123.0){\rule[-0.200pt]{0.400pt}{177.543pt}}
\put(161.0,860.0){\rule[-0.200pt]{307.870pt}{0.400pt}}
\put(40,491){\makebox(0,0){\Large{${C(t)}\over{C(0)}$}}}
\put(800,21){\makebox(0,0){\Large{$ta/\beta$}}}
\put(161.0,123.0){\rule[-0.200pt]{0.400pt}{177.543pt}}
\put(161.0,786.0){\rule[-0.200pt]{0.400pt}{3.373pt}}
\put(151.0,786.0){\rule[-0.200pt]{4.818pt}{0.400pt}}
\put(151.0,800.0){\rule[-0.200pt]{4.818pt}{0.400pt}}
\put(271.0,576.0){\rule[-0.200pt]{0.400pt}{1.927pt}}
\put(261.0,576.0){\rule[-0.200pt]{4.818pt}{0.400pt}}
\put(261.0,584.0){\rule[-0.200pt]{4.818pt}{0.400pt}}
\put(380.0,322.0){\rule[-0.200pt]{0.400pt}{1.204pt}}
\put(370.0,322.0){\rule[-0.200pt]{4.818pt}{0.400pt}}
\put(370.0,327.0){\rule[-0.200pt]{4.818pt}{0.400pt}}
\put(490.0,195.0){\rule[-0.200pt]{0.400pt}{0.723pt}}
\put(480.0,195.0){\rule[-0.200pt]{4.818pt}{0.400pt}}
\put(480.0,198.0){\rule[-0.200pt]{4.818pt}{0.400pt}}
\put(599.0,148.0){\rule[-0.200pt]{0.400pt}{0.482pt}}
\put(589.0,148.0){\rule[-0.200pt]{4.818pt}{0.400pt}}
\put(589.0,150.0){\rule[-0.200pt]{4.818pt}{0.400pt}}
\put(709.0,135.0){\rule[-0.200pt]{0.400pt}{0.482pt}}
\put(699.0,135.0){\rule[-0.200pt]{4.818pt}{0.400pt}}
\put(699.0,137.0){\rule[-0.200pt]{4.818pt}{0.400pt}}
\put(819.0,133.0){\usebox{\plotpoint}}
\put(809.0,133.0){\rule[-0.200pt]{4.818pt}{0.400pt}}
\put(809.0,134.0){\rule[-0.200pt]{4.818pt}{0.400pt}}
\put(928.0,131.0){\usebox{\plotpoint}}
\put(918.0,131.0){\rule[-0.200pt]{4.818pt}{0.400pt}}
\put(918.0,132.0){\rule[-0.200pt]{4.818pt}{0.400pt}}
\put(1038.0,127.0){\usebox{\plotpoint}}
\put(1028.0,127.0){\rule[-0.200pt]{4.818pt}{0.400pt}}
\put(1028.0,128.0){\rule[-0.200pt]{4.818pt}{0.400pt}}
\put(1147.0,124.0){\usebox{\plotpoint}}
\put(1137.0,124.0){\rule[-0.200pt]{4.818pt}{0.400pt}}
\put(161,793){\raisebox{-.8pt}{\makebox(0,0){$\Box$}}}
\put(271,580){\raisebox{-.8pt}{\makebox(0,0){$\Box$}}}
\put(380,324){\raisebox{-.8pt}{\makebox(0,0){$\Box$}}}
\put(490,197){\raisebox{-.8pt}{\makebox(0,0){$\Box$}}}
\put(599,149){\raisebox{-.8pt}{\makebox(0,0){$\Box$}}}
\put(709,136){\raisebox{-.8pt}{\makebox(0,0){$\Box$}}}
\put(819,134){\raisebox{-.8pt}{\makebox(0,0){$\Box$}}}
\put(928,131){\raisebox{-.8pt}{\makebox(0,0){$\Box$}}}
\put(1038,128){\raisebox{-.8pt}{\makebox(0,0){$\Box$}}}
\put(1147,125){\raisebox{-.8pt}{\makebox(0,0){$\Box$}}}
\put(1137.0,125.0){\rule[-0.200pt]{4.818pt}{0.400pt}}
\put(161.0,787.0){\rule[-0.200pt]{0.400pt}{2.891pt}}
\put(151.0,787.0){\rule[-0.200pt]{4.818pt}{0.400pt}}
\put(151.0,799.0){\rule[-0.200pt]{4.818pt}{0.400pt}}
\put(298.0,509.0){\rule[-0.200pt]{0.400pt}{1.686pt}}
\put(288.0,509.0){\rule[-0.200pt]{4.818pt}{0.400pt}}
\put(288.0,516.0){\rule[-0.200pt]{4.818pt}{0.400pt}}
\put(435.0,254.0){\rule[-0.200pt]{0.400pt}{0.964pt}}
\put(425.0,254.0){\rule[-0.200pt]{4.818pt}{0.400pt}}
\put(425.0,258.0){\rule[-0.200pt]{4.818pt}{0.400pt}}
\put(572.0,162.0){\rule[-0.200pt]{0.400pt}{0.723pt}}
\put(562.0,162.0){\rule[-0.200pt]{4.818pt}{0.400pt}}
\put(562.0,165.0){\rule[-0.200pt]{4.818pt}{0.400pt}}
\put(709.0,141.0){\usebox{\plotpoint}}
\put(699.0,141.0){\rule[-0.200pt]{4.818pt}{0.400pt}}
\put(699.0,142.0){\rule[-0.200pt]{4.818pt}{0.400pt}}
\put(846.0,135.0){\rule[-0.200pt]{0.400pt}{0.482pt}}
\put(836.0,135.0){\rule[-0.200pt]{4.818pt}{0.400pt}}
\put(836.0,137.0){\rule[-0.200pt]{4.818pt}{0.400pt}}
\put(983.0,130.0){\usebox{\plotpoint}}
\put(973.0,130.0){\rule[-0.200pt]{4.818pt}{0.400pt}}
\put(973.0,131.0){\rule[-0.200pt]{4.818pt}{0.400pt}}
\put(1120.0,126.0){\usebox{\plotpoint}}
\put(1110.0,126.0){\rule[-0.200pt]{4.818pt}{0.400pt}}
\put(1110.0,127.0){\rule[-0.200pt]{4.818pt}{0.400pt}}
\put(1256.0,124.0){\usebox{\plotpoint}}
\put(1246.0,124.0){\rule[-0.200pt]{4.818pt}{0.400pt}}
\put(1246.0,125.0){\rule[-0.200pt]{4.818pt}{0.400pt}}
\put(1393,124){\usebox{\plotpoint}}
\put(1383.0,124.0){\rule[-0.200pt]{4.818pt}{0.400pt}}
\put(161,793){\makebox(0,0){$\times$}}
\put(298,512){\makebox(0,0){$\times$}}
\put(435,256){\makebox(0,0){$\times$}}
\put(572,164){\makebox(0,0){$\times$}}
\put(709,142){\makebox(0,0){$\times$}}
\put(846,136){\makebox(0,0){$\times$}}
\put(983,131){\makebox(0,0){$\times$}}
\put(1120,126){\makebox(0,0){$\times$}}
\put(1256,124){\makebox(0,0){$\times$}}
\put(1393,124){\makebox(0,0){$\times$}}
\put(1383.0,124.0){\rule[-0.200pt]{4.818pt}{0.400pt}}
\put(161.0,779.0){\rule[-0.200pt]{0.400pt}{6.745pt}}
\put(151.0,779.0){\rule[-0.200pt]{4.818pt}{0.400pt}}
\put(151.0,807.0){\rule[-0.200pt]{4.818pt}{0.400pt}}
\put(271.0,571.0){\rule[-0.200pt]{0.400pt}{4.095pt}}
\put(261.0,571.0){\rule[-0.200pt]{4.818pt}{0.400pt}}
\put(261.0,588.0){\rule[-0.200pt]{4.818pt}{0.400pt}}
\put(380.0,327.0){\rule[-0.200pt]{0.400pt}{2.650pt}}
\put(370.0,327.0){\rule[-0.200pt]{4.818pt}{0.400pt}}
\put(370.0,338.0){\rule[-0.200pt]{4.818pt}{0.400pt}}
\put(490.0,207.0){\rule[-0.200pt]{0.400pt}{1.927pt}}
\put(480.0,207.0){\rule[-0.200pt]{4.818pt}{0.400pt}}
\put(480.0,215.0){\rule[-0.200pt]{4.818pt}{0.400pt}}
\put(599.0,161.0){\rule[-0.200pt]{0.400pt}{1.204pt}}
\put(589.0,161.0){\rule[-0.200pt]{4.818pt}{0.400pt}}
\put(589.0,166.0){\rule[-0.200pt]{4.818pt}{0.400pt}}
\put(709.0,146.0){\rule[-0.200pt]{0.400pt}{0.964pt}}
\put(699.0,146.0){\rule[-0.200pt]{4.818pt}{0.400pt}}
\put(699.0,150.0){\rule[-0.200pt]{4.818pt}{0.400pt}}
\put(818.0,140.0){\rule[-0.200pt]{0.400pt}{0.964pt}}
\put(808.0,140.0){\rule[-0.200pt]{4.818pt}{0.400pt}}
\put(808.0,144.0){\rule[-0.200pt]{4.818pt}{0.400pt}}
\put(928.0,135.0){\rule[-0.200pt]{0.400pt}{0.723pt}}
\put(918.0,135.0){\rule[-0.200pt]{4.818pt}{0.400pt}}
\put(918.0,138.0){\rule[-0.200pt]{4.818pt}{0.400pt}}
\put(1037.0,130.0){\rule[-0.200pt]{0.400pt}{0.723pt}}
\put(1027.0,130.0){\rule[-0.200pt]{4.818pt}{0.400pt}}
\put(1027.0,133.0){\rule[-0.200pt]{4.818pt}{0.400pt}}
\put(1147.0,127.0){\rule[-0.200pt]{0.400pt}{0.482pt}}
\put(1137.0,127.0){\rule[-0.200pt]{4.818pt}{0.400pt}}
\put(161,793){\makebox(0,0){$\triangle$}}
\put(271,579){\makebox(0,0){$\triangle$}}
\put(380,332){\makebox(0,0){$\triangle$}}
\put(490,211){\makebox(0,0){$\triangle$}}
\put(599,164){\makebox(0,0){$\triangle$}}
\put(709,148){\makebox(0,0){$\triangle$}}
\put(818,142){\makebox(0,0){$\triangle$}}
\put(928,136){\makebox(0,0){$\triangle$}}
\put(1037,132){\makebox(0,0){$\triangle$}}
\put(1147,128){\makebox(0,0){$\triangle$}}
\put(1137.0,129.0){\rule[-0.200pt]{4.818pt}{0.400pt}}
\put(161.0,778.0){\rule[-0.200pt]{0.400pt}{7.227pt}}
\put(151.0,778.0){\rule[-0.200pt]{4.818pt}{0.400pt}}
\put(151.0,808.0){\rule[-0.200pt]{4.818pt}{0.400pt}}
\put(252.0,622.0){\rule[-0.200pt]{0.400pt}{5.782pt}}
\put(242.0,622.0){\rule[-0.200pt]{4.818pt}{0.400pt}}
\put(242.0,646.0){\rule[-0.200pt]{4.818pt}{0.400pt}}
\put(344.0,401.0){\rule[-0.200pt]{0.400pt}{3.613pt}}
\put(334.0,401.0){\rule[-0.200pt]{4.818pt}{0.400pt}}
\put(334.0,416.0){\rule[-0.200pt]{4.818pt}{0.400pt}}
\put(435.0,265.0){\rule[-0.200pt]{0.400pt}{2.650pt}}
\put(425.0,265.0){\rule[-0.200pt]{4.818pt}{0.400pt}}
\put(425.0,276.0){\rule[-0.200pt]{4.818pt}{0.400pt}}
\put(526.0,198.0){\rule[-0.200pt]{0.400pt}{1.927pt}}
\put(516.0,198.0){\rule[-0.200pt]{4.818pt}{0.400pt}}
\put(516.0,206.0){\rule[-0.200pt]{4.818pt}{0.400pt}}
\put(617.0,170.0){\rule[-0.200pt]{0.400pt}{1.204pt}}
\put(607.0,170.0){\rule[-0.200pt]{4.818pt}{0.400pt}}
\put(607.0,175.0){\rule[-0.200pt]{4.818pt}{0.400pt}}
\put(709.0,157.0){\rule[-0.200pt]{0.400pt}{0.964pt}}
\put(699.0,157.0){\rule[-0.200pt]{4.818pt}{0.400pt}}
\put(699.0,161.0){\rule[-0.200pt]{4.818pt}{0.400pt}}
\put(800.0,149.0){\rule[-0.200pt]{0.400pt}{0.723pt}}
\put(790.0,149.0){\rule[-0.200pt]{4.818pt}{0.400pt}}
\put(790.0,152.0){\rule[-0.200pt]{4.818pt}{0.400pt}}
\put(891.0,141.0){\rule[-0.200pt]{0.400pt}{0.723pt}}
\put(881.0,141.0){\rule[-0.200pt]{4.818pt}{0.400pt}}
\put(881.0,144.0){\rule[-0.200pt]{4.818pt}{0.400pt}}
\put(983.0,134.0){\rule[-0.200pt]{0.400pt}{0.723pt}}
\put(973.0,134.0){\rule[-0.200pt]{4.818pt}{0.400pt}}
\put(161,793){\makebox(0,0){$\star$}}
\put(252,634){\makebox(0,0){$\star$}}
\put(344,408){\makebox(0,0){$\star$}}
\put(435,270){\makebox(0,0){$\star$}}
\put(526,202){\makebox(0,0){$\star$}}
\put(617,172){\makebox(0,0){$\star$}}
\put(709,159){\makebox(0,0){$\star$}}
\put(800,151){\makebox(0,0){$\star$}}
\put(891,143){\makebox(0,0){$\star$}}
\put(983,136){\makebox(0,0){$\star$}}
\put(973.0,137.0){\rule[-0.200pt]{4.818pt}{0.400pt}}
\end{picture}
\caption{The autocorrelator
$\langle {\bf D}\times{\bf B}({\bf x},t)\cdot{\bf D}\times{\bf B}({\bf x},0)
\rangle$ versus time
$t$ in units of $(g^2T)^{-1}$ for $\beta=8.33$ (crosses), 
$\beta=10.0$ (squares), $\beta=12.5$ (triangles), and $\beta=15.0$ (stars).}
\label{bbrr}
\end{figure}
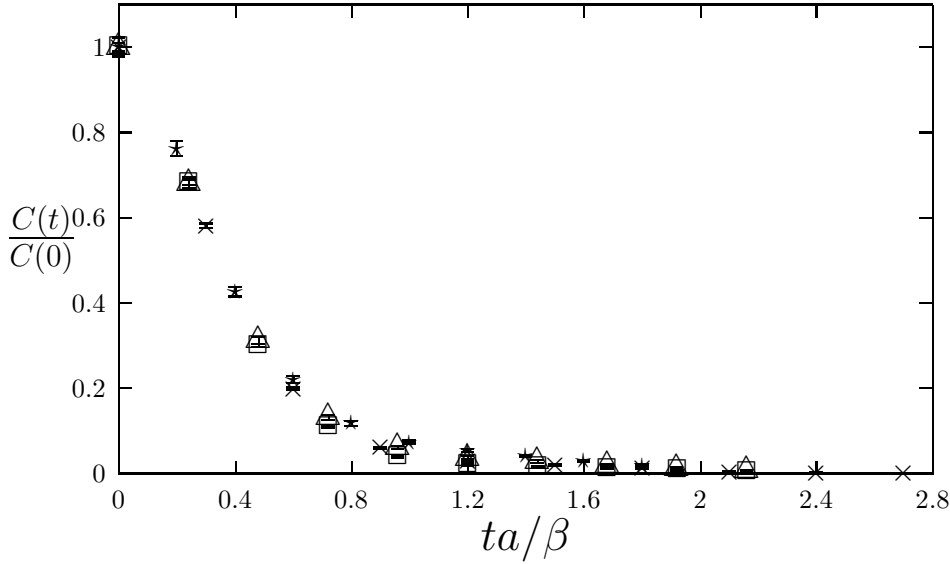
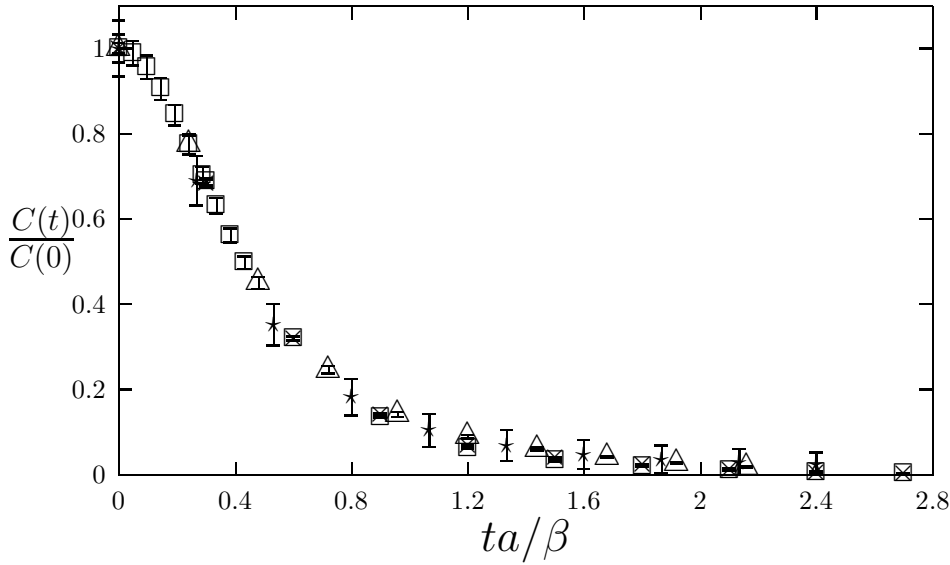
\begin{figure}[hbt]
%\input mfrr
% GNUPLOT: LaTeX picture
\setlength{\unitlength}{0.240900pt}
\ifx\plotpoint\undefined\newsavebox{\plotpoint}\fi
\sbox{\plotpoint}{\rule[-0.200pt]{0.400pt}{0.400pt}}%
\begin{picture}(1500,900)(0,0)
\font\gnuplot=cmr10 at 10pt
\gnuplot
\sbox{\plotpoint}{\rule[-0.200pt]{0.400pt}{0.400pt}}%
\put(161.0,123.0){\rule[-0.200pt]{4.818pt}{0.400pt}}
\put(141,123){\makebox(0,0)[r]{0}}
\put(1419.0,123.0){\rule[-0.200pt]{4.818pt}{0.400pt}}
\put(161.0,257.0){\rule[-0.200pt]{4.818pt}{0.400pt}}
\put(141,257){\makebox(0,0)[r]{0.2}}
\put(1419.0,257.0){\rule[-0.200pt]{4.818pt}{0.400pt}}
\put(161.0,391.0){\rule[-0.200pt]{4.818pt}{0.400pt}}
\put(141,391){\makebox(0,0)[r]{0.4}}
\put(1419.0,391.0){\rule[-0.200pt]{4.818pt}{0.400pt}}
\put(161.0,525.0){\rule[-0.200pt]{4.818pt}{0.400pt}}
\put(141,525){\makebox(0,0)[r]{0.6}}
\put(1419.0,525.0){\rule[-0.200pt]{4.818pt}{0.400pt}}
\put(161.0,659.0){\rule[-0.200pt]{4.818pt}{0.400pt}}
\put(141,659){\makebox(0,0)[r]{0.8}}
\put(1419.0,659.0){\rule[-0.200pt]{4.818pt}{0.400pt}}
\put(161.0,793.0){\rule[-0.200pt]{4.818pt}{0.400pt}}
\put(141,793){\makebox(0,0)[r]{1}}
\put(1419.0,793.0){\rule[-0.200pt]{4.818pt}{0.400pt}}
\put(161.0,123.0){\rule[-0.200pt]{0.400pt}{4.818pt}}
\put(161,82){\makebox(0,0){0}}
\put(161.0,840.0){\rule[-0.200pt]{0.400pt}{4.818pt}}
\put(344.0,123.0){\rule[-0.200pt]{0.400pt}{4.818pt}}
\put(344,82){\makebox(0,0){0.4}}
\put(344.0,840.0){\rule[-0.200pt]{0.400pt}{4.818pt}}
\put(526.0,123.0){\rule[-0.200pt]{0.400pt}{4.818pt}}
\put(526,82){\makebox(0,0){0.8}}
\put(526.0,840.0){\rule[-0.200pt]{0.400pt}{4.818pt}}
\put(709.0,123.0){\rule[-0.200pt]{0.400pt}{4.818pt}}
\put(709,82){\makebox(0,0){1.2}}
\put(709.0,840.0){\rule[-0.200pt]{0.400pt}{4.818pt}}
\put(891.0,123.0){\rule[-0.200pt]{0.400pt}{4.818pt}}
\put(891,82){\makebox(0,0){1.6}}
\put(891.0,840.0){\rule[-0.200pt]{0.400pt}{4.818pt}}
\put(1074.0,123.0){\rule[-0.200pt]{0.400pt}{4.818pt}}
\put(1074,82){\makebox(0,0){2}}
\put(1074.0,840.0){\rule[-0.200pt]{0.400pt}{4.818pt}}
\put(1256.0,123.0){\rule[-0.200pt]{0.400pt}{4.818pt}}
\put(1256,82){\makebox(0,0){2.4}}
\put(1256.0,840.0){\rule[-0.200pt]{0.400pt}{4.818pt}}
\put(1439.0,123.0){\rule[-0.200pt]{0.400pt}{4.818pt}}
\put(1439,82){\makebox(0,0){2.8}}
\put(1439.0,840.0){\rule[-0.200pt]{0.400pt}{4.818pt}}
\put(161.0,123.0){\rule[-0.200pt]{307.870pt}{0.400pt}}
\put(1439.0,123.0){\rule[-0.200pt]{0.400pt}{177.543pt}}
\put(161.0,860.0){\rule[-0.200pt]{307.870pt}{0.400pt}}
\put(40,491){\makebox(0,0){\Large{${C(t)}\over{C(0)}$}}}
\put(800,21){\makebox(0,0){\Large{$ta/\beta$}}}
\put(161.0,123.0){\rule[-0.200pt]{0.400pt}{177.543pt}}
\put(161.0,785.0){\rule[-0.200pt]{0.400pt}{3.854pt}}
\put(151.0,785.0){\rule[-0.200pt]{4.818pt}{0.400pt}}
\put(151.0,801.0){\rule[-0.200pt]{4.818pt}{0.400pt}}
\put(183.0,766.0){\rule[-0.200pt]{0.400pt}{9.395pt}}
\put(173.0,766.0){\rule[-0.200pt]{4.818pt}{0.400pt}}
\put(173.0,805.0){\rule[-0.200pt]{4.818pt}{0.400pt}}
\put(205.0,745.0){\rule[-0.200pt]{0.400pt}{8.913pt}}
\put(195.0,745.0){\rule[-0.200pt]{4.818pt}{0.400pt}}
\put(195.0,782.0){\rule[-0.200pt]{4.818pt}{0.400pt}}
\put(227.0,712.0){\rule[-0.200pt]{0.400pt}{8.431pt}}
\put(217.0,712.0){\rule[-0.200pt]{4.818pt}{0.400pt}}
\put(217.0,747.0){\rule[-0.200pt]{4.818pt}{0.400pt}}
\put(249.0,672.0){\rule[-0.200pt]{0.400pt}{7.709pt}}
\put(239.0,672.0){\rule[-0.200pt]{4.818pt}{0.400pt}}
\put(239.0,704.0){\rule[-0.200pt]{4.818pt}{0.400pt}}
\put(271.0,627.0){\rule[-0.200pt]{0.400pt}{6.986pt}}
\put(261.0,627.0){\rule[-0.200pt]{4.818pt}{0.400pt}}
\put(261.0,656.0){\rule[-0.200pt]{4.818pt}{0.400pt}}
\put(292.0,580.0){\rule[-0.200pt]{0.400pt}{6.504pt}}
\put(282.0,580.0){\rule[-0.200pt]{4.818pt}{0.400pt}}
\put(282.0,607.0){\rule[-0.200pt]{4.818pt}{0.400pt}}
\put(298.0,578.0){\rule[-0.200pt]{0.400pt}{2.650pt}}
\put(288.0,578.0){\rule[-0.200pt]{4.818pt}{0.400pt}}
\put(288.0,589.0){\rule[-0.200pt]{4.818pt}{0.400pt}}
\put(314.0,533.0){\rule[-0.200pt]{0.400pt}{6.022pt}}
\put(304.0,533.0){\rule[-0.200pt]{4.818pt}{0.400pt}}
\put(304.0,558.0){\rule[-0.200pt]{4.818pt}{0.400pt}}
\put(336.0,488.0){\rule[-0.200pt]{0.400pt}{5.300pt}}
\put(326.0,488.0){\rule[-0.200pt]{4.818pt}{0.400pt}}
\put(326.0,510.0){\rule[-0.200pt]{4.818pt}{0.400pt}}
\put(358.0,446.0){\rule[-0.200pt]{0.400pt}{4.818pt}}
\put(348.0,446.0){\rule[-0.200pt]{4.818pt}{0.400pt}}
\put(348.0,466.0){\rule[-0.200pt]{4.818pt}{0.400pt}}
\put(435.0,334.0){\rule[-0.200pt]{0.400pt}{1.445pt}}
\put(425.0,334.0){\rule[-0.200pt]{4.818pt}{0.400pt}}
\put(425.0,340.0){\rule[-0.200pt]{4.818pt}{0.400pt}}
\put(572.0,212.0){\rule[-0.200pt]{0.400pt}{0.723pt}}
\put(562.0,212.0){\rule[-0.200pt]{4.818pt}{0.400pt}}
\put(562.0,215.0){\rule[-0.200pt]{4.818pt}{0.400pt}}
\put(709.0,163.0){\rule[-0.200pt]{0.400pt}{0.482pt}}
\put(699.0,163.0){\rule[-0.200pt]{4.818pt}{0.400pt}}
\put(699.0,165.0){\rule[-0.200pt]{4.818pt}{0.400pt}}
\put(846.0,144.0){\rule[-0.200pt]{0.400pt}{0.482pt}}
\put(836.0,144.0){\rule[-0.200pt]{4.818pt}{0.400pt}}
\put(836.0,146.0){\rule[-0.200pt]{4.818pt}{0.400pt}}
\put(983.0,135.0){\usebox{\plotpoint}}
\put(973.0,135.0){\rule[-0.200pt]{4.818pt}{0.400pt}}
\put(973.0,136.0){\rule[-0.200pt]{4.818pt}{0.400pt}}
\put(1120.0,129.0){\usebox{\plotpoint}}
\put(1110.0,129.0){\rule[-0.200pt]{4.818pt}{0.400pt}}
\put(1110.0,130.0){\rule[-0.200pt]{4.818pt}{0.400pt}}
\put(1256.0,126.0){\usebox{\plotpoint}}
\put(1246.0,126.0){\rule[-0.200pt]{4.818pt}{0.400pt}}
\put(1246.0,127.0){\rule[-0.200pt]{4.818pt}{0.400pt}}
\put(1393.0,124.0){\usebox{\plotpoint}}
\put(1383.0,124.0){\rule[-0.200pt]{4.818pt}{0.400pt}}
\put(161,793){\raisebox{-.8pt}{\makebox(0,0){$\Box$}}}
\put(183,785){\raisebox{-.8pt}{\makebox(0,0){$\Box$}}}
\put(205,763){\raisebox{-.8pt}{\makebox(0,0){$\Box$}}}
\put(227,729){\raisebox{-.8pt}{\makebox(0,0){$\Box$}}}
\put(249,688){\raisebox{-.8pt}{\makebox(0,0){$\Box$}}}
\put(271,641){\raisebox{-.8pt}{\makebox(0,0){$\Box$}}}
\put(292,593){\raisebox{-.8pt}{\makebox(0,0){$\Box$}}}
\put(298,583){\raisebox{-.8pt}{\makebox(0,0){$\Box$}}}
\put(314,545){\raisebox{-.8pt}{\makebox(0,0){$\Box$}}}
\put(336,499){\raisebox{-.8pt}{\makebox(0,0){$\Box$}}}
\put(358,456){\raisebox{-.8pt}{\makebox(0,0){$\Box$}}}
\put(435,337){\raisebox{-.8pt}{\makebox(0,0){$\Box$}}}
\put(572,213){\raisebox{-.8pt}{\makebox(0,0){$\Box$}}}
\put(709,164){\raisebox{-.8pt}{\makebox(0,0){$\Box$}}}
\put(846,145){\raisebox{-.8pt}{\makebox(0,0){$\Box$}}}
\put(983,135){\raisebox{-.8pt}{\makebox(0,0){$\Box$}}}
\put(1120,130){\raisebox{-.8pt}{\makebox(0,0){$\Box$}}}
\put(1256,126){\raisebox{-.8pt}{\makebox(0,0){$\Box$}}}
\put(1393,125){\raisebox{-.8pt}{\makebox(0,0){$\Box$}}}
\put(1383.0,125.0){\rule[-0.200pt]{4.818pt}{0.400pt}}
\put(161.0,784.0){\rule[-0.200pt]{0.400pt}{4.336pt}}
\put(151.0,784.0){\rule[-0.200pt]{4.818pt}{0.400pt}}
\put(151.0,802.0){\rule[-0.200pt]{4.818pt}{0.400pt}}
\put(298.0,575.0){\rule[-0.200pt]{0.400pt}{2.891pt}}
\put(288.0,575.0){\rule[-0.200pt]{4.818pt}{0.400pt}}
\put(288.0,587.0){\rule[-0.200pt]{4.818pt}{0.400pt}}
\put(435.0,334.0){\rule[-0.200pt]{0.400pt}{1.686pt}}
\put(425.0,334.0){\rule[-0.200pt]{4.818pt}{0.400pt}}
\put(425.0,341.0){\rule[-0.200pt]{4.818pt}{0.400pt}}
\put(572.0,216.0){\rule[-0.200pt]{0.400pt}{0.964pt}}
\put(562.0,216.0){\rule[-0.200pt]{4.818pt}{0.400pt}}
\put(562.0,220.0){\rule[-0.200pt]{4.818pt}{0.400pt}}
\put(709.0,168.0){\rule[-0.200pt]{0.400pt}{0.723pt}}
\put(699.0,168.0){\rule[-0.200pt]{4.818pt}{0.400pt}}
\put(699.0,171.0){\rule[-0.200pt]{4.818pt}{0.400pt}}
\put(846.0,149.0){\rule[-0.200pt]{0.400pt}{0.482pt}}
\put(836.0,149.0){\rule[-0.200pt]{4.818pt}{0.400pt}}
\put(836.0,151.0){\rule[-0.200pt]{4.818pt}{0.400pt}}
\put(983.0,138.0){\rule[-0.200pt]{0.400pt}{0.482pt}}
\put(973.0,138.0){\rule[-0.200pt]{4.818pt}{0.400pt}}
\put(973.0,140.0){\rule[-0.200pt]{4.818pt}{0.400pt}}
\put(1120.0,131.0){\rule[-0.200pt]{0.400pt}{0.482pt}}
\put(1110.0,131.0){\rule[-0.200pt]{4.818pt}{0.400pt}}
\put(1110.0,133.0){\rule[-0.200pt]{4.818pt}{0.400pt}}
\put(1256.0,127.0){\usebox{\plotpoint}}
\put(1246.0,127.0){\rule[-0.200pt]{4.818pt}{0.400pt}}
\put(1246.0,128.0){\rule[-0.200pt]{4.818pt}{0.400pt}}
\put(1393.0,125.0){\usebox{\plotpoint}}
\put(1383.0,125.0){\rule[-0.200pt]{4.818pt}{0.400pt}}
\put(161,793){\makebox(0,0){$\times$}}
\put(298,581){\makebox(0,0){$\times$}}
\put(435,337){\makebox(0,0){$\times$}}
\put(572,218){\makebox(0,0){$\times$}}
\put(709,170){\makebox(0,0){$\times$}}
\put(846,150){\makebox(0,0){$\times$}}
\put(983,139){\makebox(0,0){$\times$}}
\put(1120,132){\makebox(0,0){$\times$}}
\put(1256,128){\makebox(0,0){$\times$}}
\put(1393,126){\makebox(0,0){$\times$}}
\put(1383.0,126.0){\rule[-0.200pt]{4.818pt}{0.400pt}}
\put(161.0,771.0){\rule[-0.200pt]{0.400pt}{10.600pt}}
\put(151.0,771.0){\rule[-0.200pt]{4.818pt}{0.400pt}}
\put(151.0,815.0){\rule[-0.200pt]{4.818pt}{0.400pt}}
\put(271.0,625.0){\rule[-0.200pt]{0.400pt}{7.950pt}}
\put(261.0,625.0){\rule[-0.200pt]{4.818pt}{0.400pt}}
\put(261.0,658.0){\rule[-0.200pt]{4.818pt}{0.400pt}}
\put(380.0,415.0){\rule[-0.200pt]{0.400pt}{4.577pt}}
\put(370.0,415.0){\rule[-0.200pt]{4.818pt}{0.400pt}}
\put(370.0,434.0){\rule[-0.200pt]{4.818pt}{0.400pt}}
\put(490.0,282.0){\rule[-0.200pt]{0.400pt}{2.891pt}}
\put(480.0,282.0){\rule[-0.200pt]{4.818pt}{0.400pt}}
\put(480.0,294.0){\rule[-0.200pt]{4.818pt}{0.400pt}}
\put(599.0,214.0){\rule[-0.200pt]{0.400pt}{1.927pt}}
\put(589.0,214.0){\rule[-0.200pt]{4.818pt}{0.400pt}}
\put(589.0,222.0){\rule[-0.200pt]{4.818pt}{0.400pt}}
\put(709.0,180.0){\rule[-0.200pt]{0.400pt}{1.445pt}}
\put(699.0,180.0){\rule[-0.200pt]{4.818pt}{0.400pt}}
\put(699.0,186.0){\rule[-0.200pt]{4.818pt}{0.400pt}}
\put(818.0,161.0){\rule[-0.200pt]{0.400pt}{1.204pt}}
\put(808.0,161.0){\rule[-0.200pt]{4.818pt}{0.400pt}}
\put(808.0,166.0){\rule[-0.200pt]{4.818pt}{0.400pt}}
\put(928.0,149.0){\rule[-0.200pt]{0.400pt}{0.964pt}}
\put(918.0,149.0){\rule[-0.200pt]{4.818pt}{0.400pt}}
\put(918.0,153.0){\rule[-0.200pt]{4.818pt}{0.400pt}}
\put(1037.0,140.0){\rule[-0.200pt]{0.400pt}{0.723pt}}
\put(1027.0,140.0){\rule[-0.200pt]{4.818pt}{0.400pt}}
\put(1027.0,143.0){\rule[-0.200pt]{4.818pt}{0.400pt}}
\put(1147.0,134.0){\rule[-0.200pt]{0.400pt}{0.482pt}}
\put(1137.0,134.0){\rule[-0.200pt]{4.818pt}{0.400pt}}
\put(161,793){\makebox(0,0){$\triangle$}}
\put(271,642){\makebox(0,0){$\triangle$}}
\put(380,425){\makebox(0,0){$\triangle$}}
\put(490,288){\makebox(0,0){$\triangle$}}
\put(599,218){\makebox(0,0){$\triangle$}}
\put(709,183){\makebox(0,0){$\triangle$}}
\put(818,164){\makebox(0,0){$\triangle$}}
\put(928,151){\makebox(0,0){$\triangle$}}
\put(1037,141){\makebox(0,0){$\triangle$}}
\put(1147,135){\makebox(0,0){$\triangle$}}
\put(1137.0,136.0){\rule[-0.200pt]{4.818pt}{0.400pt}}
\put(161.0,749.0){\rule[-0.200pt]{0.400pt}{21.199pt}}
\put(151.0,749.0){\rule[-0.200pt]{4.818pt}{0.400pt}}
\put(151.0,837.0){\rule[-0.200pt]{4.818pt}{0.400pt}}
\put(283.0,546.0){\rule[-0.200pt]{0.400pt}{18.790pt}}
\put(273.0,546.0){\rule[-0.200pt]{4.818pt}{0.400pt}}
\put(273.0,624.0){\rule[-0.200pt]{4.818pt}{0.400pt}}
\put(404.0,326.0){\rule[-0.200pt]{0.400pt}{15.658pt}}
\put(394.0,326.0){\rule[-0.200pt]{4.818pt}{0.400pt}}
\put(394.0,391.0){\rule[-0.200pt]{4.818pt}{0.400pt}}
\put(526.0,216.0){\rule[-0.200pt]{0.400pt}{13.972pt}}
\put(516.0,216.0){\rule[-0.200pt]{4.818pt}{0.400pt}}
\put(516.0,274.0){\rule[-0.200pt]{4.818pt}{0.400pt}}
\put(648.0,167.0){\rule[-0.200pt]{0.400pt}{12.527pt}}
\put(638.0,167.0){\rule[-0.200pt]{4.818pt}{0.400pt}}
\put(638.0,219.0){\rule[-0.200pt]{4.818pt}{0.400pt}}
\put(770.0,145.0){\rule[-0.200pt]{0.400pt}{11.563pt}}
\put(760.0,145.0){\rule[-0.200pt]{4.818pt}{0.400pt}}
\put(760.0,193.0){\rule[-0.200pt]{4.818pt}{0.400pt}}
\put(891.0,132.0){\rule[-0.200pt]{0.400pt}{11.081pt}}
\put(881.0,132.0){\rule[-0.200pt]{4.818pt}{0.400pt}}
\put(881.0,178.0){\rule[-0.200pt]{4.818pt}{0.400pt}}
\put(1013.0,125.0){\rule[-0.200pt]{0.400pt}{10.600pt}}
\put(1003.0,125.0){\rule[-0.200pt]{4.818pt}{0.400pt}}
\put(1003.0,169.0){\rule[-0.200pt]{4.818pt}{0.400pt}}
\put(1135.0,123.0){\rule[-0.200pt]{0.400pt}{9.636pt}}
\put(1125.0,123.0){\rule[-0.200pt]{4.818pt}{0.400pt}}
\put(1125.0,163.0){\rule[-0.200pt]{4.818pt}{0.400pt}}
\put(1256.0,123.0){\rule[-0.200pt]{0.400pt}{8.431pt}}
\put(1246.0,123.0){\rule[-0.200pt]{4.818pt}{0.400pt}}
\put(161,793){\makebox(0,0){$\star$}}
\put(283,585){\makebox(0,0){$\star$}}
\put(404,359){\makebox(0,0){$\star$}}
\put(526,245){\makebox(0,0){$\star$}}
\put(648,193){\makebox(0,0){$\star$}}
\put(770,169){\makebox(0,0){$\star$}}
\put(891,155){\makebox(0,0){$\star$}}
\put(1013,147){\makebox(0,0){$\star$}}
\put(1135,142){\makebox(0,0){$\star$}}
\put(1256,138){\makebox(0,0){$\star$}}
\put(1246.0,158.0){\rule[-0.200pt]{4.818pt}{0.400pt}}
\end{picture}
\caption{The autocorrelator
$\langle {\bf B}({\bf x},t)\cdot{\bf B}({\bf x},0)
\rangle$ versus time
$t$ in units of $(g^2T)^{-1}$ for $\beta=8.33$ (crosses), 
$\beta=10.0$ (squares), $\beta=12.5$ (triangles), and $\beta=15.0$ (stars).}
\label{mfrr}
\end{figure}
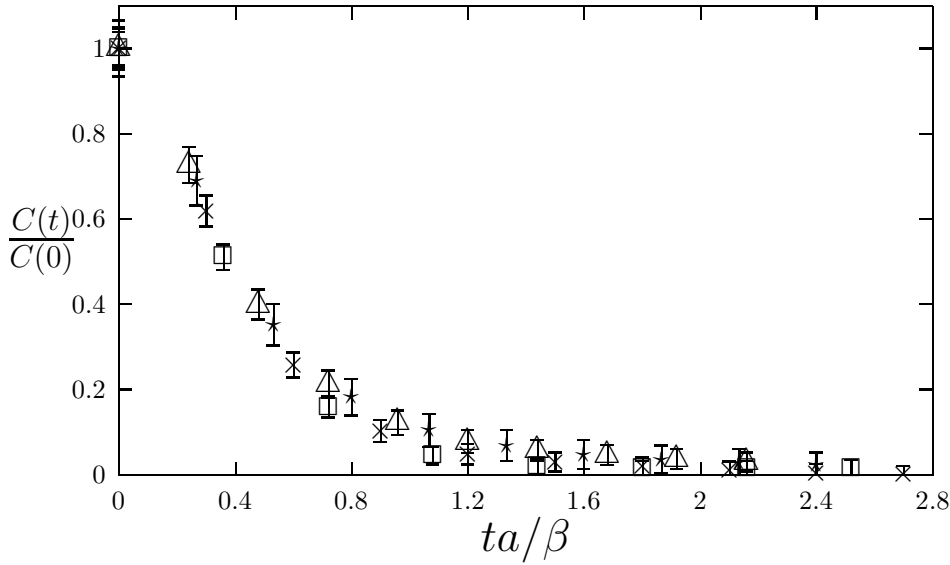
\begin{figure}[hbt]
%\input plcrr
% GNUPLOT: LaTeX picture
\setlength{\unitlength}{0.240900pt}
\ifx\plotpoint\undefined\newsavebox{\plotpoint}\fi
\sbox{\plotpoint}{\rule[-0.200pt]{0.400pt}{0.400pt}}%
\begin{picture}(1500,900)(0,0)
\font\gnuplot=cmr10 at 10pt
\gnuplot
\sbox{\plotpoint}{\rule[-0.200pt]{0.400pt}{0.400pt}}%
\put(161.0,123.0){\rule[-0.200pt]{4.818pt}{0.400pt}}
\put(141,123){\makebox(0,0)[r]{0}}
\put(1419.0,123.0){\rule[-0.200pt]{4.818pt}{0.400pt}}
\put(161.0,257.0){\rule[-0.200pt]{4.818pt}{0.400pt}}
\put(141,257){\makebox(0,0)[r]{0.2}}
\put(1419.0,257.0){\rule[-0.200pt]{4.818pt}{0.400pt}}
\put(161.0,391.0){\rule[-0.200pt]{4.818pt}{0.400pt}}
\put(141,391){\makebox(0,0)[r]{0.4}}
\put(1419.0,391.0){\rule[-0.200pt]{4.818pt}{0.400pt}}
\put(161.0,525.0){\rule[-0.200pt]{4.818pt}{0.400pt}}
\put(141,525){\makebox(0,0)[r]{0.6}}
\put(1419.0,525.0){\rule[-0.200pt]{4.818pt}{0.400pt}}
\put(161.0,659.0){\rule[-0.200pt]{4.818pt}{0.400pt}}
\put(141,659){\makebox(0,0)[r]{0.8}}
\put(1419.0,659.0){\rule[-0.200pt]{4.818pt}{0.400pt}}
\put(161.0,793.0){\rule[-0.200pt]{4.818pt}{0.400pt}}
\put(141,793){\makebox(0,0)[r]{1}}
\put(1419.0,793.0){\rule[-0.200pt]{4.818pt}{0.400pt}}
\put(161.0,123.0){\rule[-0.200pt]{0.400pt}{4.818pt}}
\put(161,82){\makebox(0,0){0}}
\put(161.0,840.0){\rule[-0.200pt]{0.400pt}{4.818pt}}
\put(344.0,123.0){\rule[-0.200pt]{0.400pt}{4.818pt}}
\put(344,82){\makebox(0,0){0.4}}
\put(344.0,840.0){\rule[-0.200pt]{0.400pt}{4.818pt}}
\put(526.0,123.0){\rule[-0.200pt]{0.400pt}{4.818pt}}
\put(526,82){\makebox(0,0){0.8}}
\put(526.0,840.0){\rule[-0.200pt]{0.400pt}{4.818pt}}
\put(709.0,123.0){\rule[-0.200pt]{0.400pt}{4.818pt}}
\put(709,82){\makebox(0,0){1.2}}
\put(709.0,840.0){\rule[-0.200pt]{0.400pt}{4.818pt}}
\put(891.0,123.0){\rule[-0.200pt]{0.400pt}{4.818pt}}
\put(891,82){\makebox(0,0){1.6}}
\put(891.0,840.0){\rule[-0.200pt]{0.400pt}{4.818pt}}
\put(1074.0,123.0){\rule[-0.200pt]{0.400pt}{4.818pt}}
\put(1074,82){\makebox(0,0){2}}
\put(1074.0,840.0){\rule[-0.200pt]{0.400pt}{4.818pt}}
\put(1256.0,123.0){\rule[-0.200pt]{0.400pt}{4.818pt}}
\put(1256,82){\makebox(0,0){2.4}}
\put(1256.0,840.0){\rule[-0.200pt]{0.400pt}{4.818pt}}
\put(1439.0,123.0){\rule[-0.200pt]{0.400pt}{4.818pt}}
\put(1439,82){\makebox(0,0){2.8}}
\put(1439.0,840.0){\rule[-0.200pt]{0.400pt}{4.818pt}}
\put(161.0,123.0){\rule[-0.200pt]{307.870pt}{0.400pt}}
\put(1439.0,123.0){\rule[-0.200pt]{0.400pt}{177.543pt}}
\put(161.0,860.0){\rule[-0.200pt]{307.870pt}{0.400pt}}
\put(40,491){\makebox(0,0){\Large{${C(t)}\over{C(0)}$}}}
\put(800,21){\makebox(0,0){\Large{$ta/\beta$}}}
\put(161.0,123.0){\rule[-0.200pt]{0.400pt}{177.543pt}}
\put(161.0,767.0){\rule[-0.200pt]{0.400pt}{12.527pt}}
\put(151.0,767.0){\rule[-0.200pt]{4.818pt}{0.400pt}}
\put(151.0,819.0){\rule[-0.200pt]{4.818pt}{0.400pt}}
\put(325.0,445.0){\rule[-0.200pt]{0.400pt}{9.636pt}}
\put(315.0,445.0){\rule[-0.200pt]{4.818pt}{0.400pt}}
\put(315.0,485.0){\rule[-0.200pt]{4.818pt}{0.400pt}}
\put(490.0,213.0){\rule[-0.200pt]{0.400pt}{7.468pt}}
\put(480.0,213.0){\rule[-0.200pt]{4.818pt}{0.400pt}}
\put(480.0,244.0){\rule[-0.200pt]{4.818pt}{0.400pt}}
\put(654.0,139.0){\rule[-0.200pt]{0.400pt}{6.745pt}}
\put(644.0,139.0){\rule[-0.200pt]{4.818pt}{0.400pt}}
\put(644.0,167.0){\rule[-0.200pt]{4.818pt}{0.400pt}}
\put(818.0,123.0){\rule[-0.200pt]{0.400pt}{6.022pt}}
\put(808.0,123.0){\rule[-0.200pt]{4.818pt}{0.400pt}}
\put(808.0,148.0){\rule[-0.200pt]{4.818pt}{0.400pt}}
\put(983.0,123.0){\rule[-0.200pt]{0.400pt}{5.300pt}}
\put(973.0,123.0){\rule[-0.200pt]{4.818pt}{0.400pt}}
\put(973.0,145.0){\rule[-0.200pt]{4.818pt}{0.400pt}}
\put(1147.0,123.0){\rule[-0.200pt]{0.400pt}{5.300pt}}
\put(1137.0,123.0){\rule[-0.200pt]{4.818pt}{0.400pt}}
\put(1137.0,145.0){\rule[-0.200pt]{4.818pt}{0.400pt}}
\put(1311.0,123.0){\rule[-0.200pt]{0.400pt}{5.541pt}}
\put(1301.0,123.0){\rule[-0.200pt]{4.818pt}{0.400pt}}
\put(161,793){\raisebox{-.8pt}{\makebox(0,0){$\Box$}}}
\put(325,465){\raisebox{-.8pt}{\makebox(0,0){$\Box$}}}
\put(490,228){\raisebox{-.8pt}{\makebox(0,0){$\Box$}}}
\put(654,153){\raisebox{-.8pt}{\makebox(0,0){$\Box$}}}
\put(818,135){\raisebox{-.8pt}{\makebox(0,0){$\Box$}}}
\put(983,132){\raisebox{-.8pt}{\makebox(0,0){$\Box$}}}
\put(1147,132){\raisebox{-.8pt}{\makebox(0,0){$\Box$}}}
\put(1311,133){\raisebox{-.8pt}{\makebox(0,0){$\Box$}}}
\put(1301.0,146.0){\rule[-0.200pt]{4.818pt}{0.400pt}}
\put(161.0,763.0){\rule[-0.200pt]{0.400pt}{14.454pt}}
\put(151.0,763.0){\rule[-0.200pt]{4.818pt}{0.400pt}}
\put(151.0,823.0){\rule[-0.200pt]{4.818pt}{0.400pt}}
\put(298.0,513.0){\rule[-0.200pt]{0.400pt}{11.804pt}}
\put(288.0,513.0){\rule[-0.200pt]{4.818pt}{0.400pt}}
\put(288.0,562.0){\rule[-0.200pt]{4.818pt}{0.400pt}}
\put(435.0,276.0){\rule[-0.200pt]{0.400pt}{9.395pt}}
\put(425.0,276.0){\rule[-0.200pt]{4.818pt}{0.400pt}}
\put(425.0,315.0){\rule[-0.200pt]{4.818pt}{0.400pt}}
\put(572.0,175.0){\rule[-0.200pt]{0.400pt}{8.191pt}}
\put(562.0,175.0){\rule[-0.200pt]{4.818pt}{0.400pt}}
\put(562.0,209.0){\rule[-0.200pt]{4.818pt}{0.400pt}}
\put(709.0,139.0){\rule[-0.200pt]{0.400pt}{7.709pt}}
\put(699.0,139.0){\rule[-0.200pt]{4.818pt}{0.400pt}}
\put(699.0,171.0){\rule[-0.200pt]{4.818pt}{0.400pt}}
\put(846.0,128.0){\rule[-0.200pt]{0.400pt}{7.227pt}}
\put(836.0,128.0){\rule[-0.200pt]{4.818pt}{0.400pt}}
\put(836.0,158.0){\rule[-0.200pt]{4.818pt}{0.400pt}}
\put(983.0,123.0){\rule[-0.200pt]{0.400pt}{6.504pt}}
\put(973.0,123.0){\rule[-0.200pt]{4.818pt}{0.400pt}}
\put(973.0,150.0){\rule[-0.200pt]{4.818pt}{0.400pt}}
\put(1120.0,123.0){\rule[-0.200pt]{0.400pt}{5.059pt}}
\put(1110.0,123.0){\rule[-0.200pt]{4.818pt}{0.400pt}}
\put(1110.0,144.0){\rule[-0.200pt]{4.818pt}{0.400pt}}
\put(1256.0,123.0){\rule[-0.200pt]{0.400pt}{4.095pt}}
\put(1246.0,123.0){\rule[-0.200pt]{4.818pt}{0.400pt}}
\put(1246.0,140.0){\rule[-0.200pt]{4.818pt}{0.400pt}}
\put(1393.0,123.0){\rule[-0.200pt]{0.400pt}{3.373pt}}
\put(1383.0,123.0){\rule[-0.200pt]{4.818pt}{0.400pt}}
\put(161,793){\makebox(0,0){$\times$}}
\put(298,537){\makebox(0,0){$\times$}}
\put(435,296){\makebox(0,0){$\times$}}
\put(572,192){\makebox(0,0){$\times$}}
\put(709,155){\makebox(0,0){$\times$}}
\put(846,143){\makebox(0,0){$\times$}}
\put(983,136){\makebox(0,0){$\times$}}
\put(1120,130){\makebox(0,0){$\times$}}
\put(1256,126){\makebox(0,0){$\times$}}
\put(1393,124){\makebox(0,0){$\times$}}
\put(1383.0,137.0){\rule[-0.200pt]{4.818pt}{0.400pt}}
\put(161.0,760.0){\rule[-0.200pt]{0.400pt}{15.899pt}}
\put(151.0,760.0){\rule[-0.200pt]{4.818pt}{0.400pt}}
\put(151.0,826.0){\rule[-0.200pt]{4.818pt}{0.400pt}}
\put(271.0,582.0){\rule[-0.200pt]{0.400pt}{13.490pt}}
\put(261.0,582.0){\rule[-0.200pt]{4.818pt}{0.400pt}}
\put(261.0,638.0){\rule[-0.200pt]{4.818pt}{0.400pt}}
\put(380.0,367.0){\rule[-0.200pt]{0.400pt}{11.322pt}}
\put(370.0,367.0){\rule[-0.200pt]{4.818pt}{0.400pt}}
\put(370.0,414.0){\rule[-0.200pt]{4.818pt}{0.400pt}}
\put(490.0,246.0){\rule[-0.200pt]{0.400pt}{9.877pt}}
\put(480.0,246.0){\rule[-0.200pt]{4.818pt}{0.400pt}}
\put(480.0,287.0){\rule[-0.200pt]{4.818pt}{0.400pt}}
\put(599.0,186.0){\rule[-0.200pt]{0.400pt}{9.154pt}}
\put(589.0,186.0){\rule[-0.200pt]{4.818pt}{0.400pt}}
\put(589.0,224.0){\rule[-0.200pt]{4.818pt}{0.400pt}}
\put(709.0,157.0){\rule[-0.200pt]{0.400pt}{8.672pt}}
\put(699.0,157.0){\rule[-0.200pt]{4.818pt}{0.400pt}}
\put(699.0,193.0){\rule[-0.200pt]{4.818pt}{0.400pt}}
\put(818.0,145.0){\rule[-0.200pt]{0.400pt}{7.950pt}}
\put(808.0,145.0){\rule[-0.200pt]{4.818pt}{0.400pt}}
\put(808.0,178.0){\rule[-0.200pt]{4.818pt}{0.400pt}}
\put(928.0,138.0){\rule[-0.200pt]{0.400pt}{7.709pt}}
\put(918.0,138.0){\rule[-0.200pt]{4.818pt}{0.400pt}}
\put(918.0,170.0){\rule[-0.200pt]{4.818pt}{0.400pt}}
\put(1037.0,132.0){\rule[-0.200pt]{0.400pt}{7.468pt}}
\put(1027.0,132.0){\rule[-0.200pt]{4.818pt}{0.400pt}}
\put(1027.0,163.0){\rule[-0.200pt]{4.818pt}{0.400pt}}
\put(1147.0,128.0){\rule[-0.200pt]{0.400pt}{7.227pt}}
\put(1137.0,128.0){\rule[-0.200pt]{4.818pt}{0.400pt}}
\put(161,793){\makebox(0,0){$\triangle$}}
\put(271,610){\makebox(0,0){$\triangle$}}
\put(380,390){\makebox(0,0){$\triangle$}}
\put(490,266){\makebox(0,0){$\triangle$}}
\put(599,205){\makebox(0,0){$\triangle$}}
\put(709,175){\makebox(0,0){$\triangle$}}
\put(818,162){\makebox(0,0){$\triangle$}}
\put(928,154){\makebox(0,0){$\triangle$}}
\put(1037,147){\makebox(0,0){$\triangle$}}
\put(1147,143){\makebox(0,0){$\triangle$}}
\put(1137.0,158.0){\rule[-0.200pt]{4.818pt}{0.400pt}}
\put(161.0,749.0){\rule[-0.200pt]{0.400pt}{21.199pt}}
\put(151.0,749.0){\rule[-0.200pt]{4.818pt}{0.400pt}}
\put(151.0,837.0){\rule[-0.200pt]{4.818pt}{0.400pt}}
\put(283.0,546.0){\rule[-0.200pt]{0.400pt}{18.790pt}}
\put(273.0,546.0){\rule[-0.200pt]{4.818pt}{0.400pt}}
\put(273.0,624.0){\rule[-0.200pt]{4.818pt}{0.400pt}}
\put(404.0,326.0){\rule[-0.200pt]{0.400pt}{15.658pt}}
\put(394.0,326.0){\rule[-0.200pt]{4.818pt}{0.400pt}}
\put(394.0,391.0){\rule[-0.200pt]{4.818pt}{0.400pt}}
\put(526.0,216.0){\rule[-0.200pt]{0.400pt}{13.972pt}}
\put(516.0,216.0){\rule[-0.200pt]{4.818pt}{0.400pt}}
\put(516.0,274.0){\rule[-0.200pt]{4.818pt}{0.400pt}}
\put(648.0,167.0){\rule[-0.200pt]{0.400pt}{12.527pt}}
\put(638.0,167.0){\rule[-0.200pt]{4.818pt}{0.400pt}}
\put(638.0,219.0){\rule[-0.200pt]{4.818pt}{0.400pt}}
\put(770.0,145.0){\rule[-0.200pt]{0.400pt}{11.563pt}}
\put(760.0,145.0){\rule[-0.200pt]{4.818pt}{0.400pt}}
\put(760.0,193.0){\rule[-0.200pt]{4.818pt}{0.400pt}}
\put(891.0,132.0){\rule[-0.200pt]{0.400pt}{11.081pt}}
\put(881.0,132.0){\rule[-0.200pt]{4.818pt}{0.400pt}}
\put(881.0,178.0){\rule[-0.200pt]{4.818pt}{0.400pt}}
\put(1013.0,125.0){\rule[-0.200pt]{0.400pt}{10.600pt}}
\put(1003.0,125.0){\rule[-0.200pt]{4.818pt}{0.400pt}}
\put(1003.0,169.0){\rule[-0.200pt]{4.818pt}{0.400pt}}
\put(1135.0,123.0){\rule[-0.200pt]{0.400pt}{9.636pt}}
\put(1125.0,123.0){\rule[-0.200pt]{4.818pt}{0.400pt}}
\put(1125.0,163.0){\rule[-0.200pt]{4.818pt}{0.400pt}}
\put(1256.0,123.0){\rule[-0.200pt]{0.400pt}{8.431pt}}
\put(1246.0,123.0){\rule[-0.200pt]{4.818pt}{0.400pt}}
\put(161,793){\makebox(0,0){$\star$}}
\put(283,585){\makebox(0,0){$\star$}}
\put(404,359){\makebox(0,0){$\star$}}
\put(526,245){\makebox(0,0){$\star$}}
\put(648,193){\makebox(0,0){$\star$}}
\put(770,169){\makebox(0,0){$\star$}}
\put(891,155){\makebox(0,0){$\star$}}
\put(1013,147){\makebox(0,0){$\star$}}
\put(1135,142){\makebox(0,0){$\star$}}
\put(1256,138){\makebox(0,0){$\star$}}
\put(1246.0,158.0){\rule[-0.200pt]{4.818pt}{0.400pt}}
\end{picture}
\caption{The connected autocorrelator
$\langle B_i^2({\bf x},t)B_i^2({\bf x},0)\rangle-
\langle B_i^2({\bf x},t)\rangle\langle B_i^2({\bf x},0)\rangle$ versus time
$t$ in units of $(g^2T)^{-1}$ for $\beta=8.33$ (crosses), 
$\beta=10.0$ (squares), $\beta=12.5$ (triangles), and $\beta=15.0$ (stars).}
\label{plcrr}
\end{figure}

\begin{figure}[hbt]
%\input tauinv
% GNUPLOT: LaTeX picture
\setlength{\unitlength}{0.240900pt}
\ifx\plotpoint\undefined\newsavebox{\plotpoint}\fi
\sbox{\plotpoint}{\rule[-0.200pt]{0.400pt}{0.400pt}}%
\begin{picture}(1500,900)(0,0)
\font\gnuplot=cmr10 at 10pt
\gnuplot
\sbox{\plotpoint}{\rule[-0.200pt]{0.400pt}{0.400pt}}%
\put(161.0,123.0){\rule[-0.200pt]{4.818pt}{0.400pt}}
\put(141,123){\makebox(0,0)[r]{0}}
\put(1419.0,123.0){\rule[-0.200pt]{4.818pt}{0.400pt}}
\put(161.0,246.0){\rule[-0.200pt]{4.818pt}{0.400pt}}
\put(141,246){\makebox(0,0)[r]{0.5}}
\put(1419.0,246.0){\rule[-0.200pt]{4.818pt}{0.400pt}}
\put(161.0,369.0){\rule[-0.200pt]{4.818pt}{0.400pt}}
\put(141,369){\makebox(0,0)[r]{1}}
\put(1419.0,369.0){\rule[-0.200pt]{4.818pt}{0.400pt}}
\put(161.0,492.0){\rule[-0.200pt]{4.818pt}{0.400pt}}
\put(141,492){\makebox(0,0)[r]{1.5}}
\put(1419.0,492.0){\rule[-0.200pt]{4.818pt}{0.400pt}}
\put(161.0,614.0){\rule[-0.200pt]{4.818pt}{0.400pt}}
\put(141,614){\makebox(0,0)[r]{2}}
\put(1419.0,614.0){\rule[-0.200pt]{4.818pt}{0.400pt}}
\put(161.0,737.0){\rule[-0.200pt]{4.818pt}{0.400pt}}
\put(141,737){\makebox(0,0)[r]{2.5}}
\put(1419.0,737.0){\rule[-0.200pt]{4.818pt}{0.400pt}}
\put(161.0,860.0){\rule[-0.200pt]{4.818pt}{0.400pt}}
\put(141,860){\makebox(0,0)[r]{3}}
\put(1419.0,860.0){\rule[-0.200pt]{4.818pt}{0.400pt}}
\put(161.0,123.0){\rule[-0.200pt]{0.400pt}{4.818pt}}
\put(161,82){\makebox(0,0){0}}
\put(161.0,840.0){\rule[-0.200pt]{0.400pt}{4.818pt}}
\put(417.0,123.0){\rule[-0.200pt]{0.400pt}{4.818pt}}
\put(417,82){\makebox(0,0){0.03}}
\put(417.0,840.0){\rule[-0.200pt]{0.400pt}{4.818pt}}
\put(672.0,123.0){\rule[-0.200pt]{0.400pt}{4.818pt}}
\put(672,82){\makebox(0,0){0.06}}
\put(672.0,840.0){\rule[-0.200pt]{0.400pt}{4.818pt}}
\put(928.0,123.0){\rule[-0.200pt]{0.400pt}{4.818pt}}
\put(928,82){\makebox(0,0){0.09}}
\put(928.0,840.0){\rule[-0.200pt]{0.400pt}{4.818pt}}
\put(1183.0,123.0){\rule[-0.200pt]{0.400pt}{4.818pt}}
\put(1183,82){\makebox(0,0){0.12}}
\put(1183.0,840.0){\rule[-0.200pt]{0.400pt}{4.818pt}}
\put(1439.0,123.0){\rule[-0.200pt]{0.400pt}{4.818pt}}
\put(1439,82){\makebox(0,0){0.15}}
\put(1439.0,840.0){\rule[-0.200pt]{0.400pt}{4.818pt}}
\put(161.0,123.0){\rule[-0.200pt]{307.870pt}{0.400pt}}
\put(1439.0,123.0){\rule[-0.200pt]{0.400pt}{177.543pt}}
\put(161.0,860.0){\rule[-0.200pt]{307.870pt}{0.400pt}}
\put(40,491){\makebox(0,0){\Large{$\beta\over{at_{\int}}$}}}
\put(800,21){\makebox(0,0){\Large{$1/\beta$}}}
\put(161.0,123.0){\rule[-0.200pt]{0.400pt}{177.543pt}}
\put(1183.0,573.0){\rule[-0.200pt]{0.400pt}{38.785pt}}
\put(1173.0,573.0){\rule[-0.200pt]{4.818pt}{0.400pt}}
\put(1173.0,734.0){\rule[-0.200pt]{4.818pt}{0.400pt}}
\put(1013.0,562.0){\rule[-0.200pt]{0.400pt}{36.376pt}}
\put(1003.0,562.0){\rule[-0.200pt]{4.818pt}{0.400pt}}
\put(1003.0,713.0){\rule[-0.200pt]{4.818pt}{0.400pt}}
\put(843.0,532.0){\rule[-0.200pt]{0.400pt}{27.703pt}}
\put(833.0,532.0){\rule[-0.200pt]{4.818pt}{0.400pt}}
\put(833.0,647.0){\rule[-0.200pt]{4.818pt}{0.400pt}}
\put(729.0,495.0){\rule[-0.200pt]{0.400pt}{40.471pt}}
\put(719.0,495.0){\rule[-0.200pt]{4.818pt}{0.400pt}}
\put(1183,653){\raisebox{-.8pt}{\makebox(0,0){$\Box$}}}
\put(1013,638){\raisebox{-.8pt}{\makebox(0,0){$\Box$}}}
\put(843,589){\raisebox{-.8pt}{\makebox(0,0){$\Box$}}}
\put(729,579){\raisebox{-.8pt}{\makebox(0,0){$\Box$}}}
\put(719.0,663.0){\rule[-0.200pt]{4.818pt}{0.400pt}}
\put(1013.0,570.0){\rule[-0.200pt]{0.400pt}{3.613pt}}
\put(1003.0,570.0){\rule[-0.200pt]{4.818pt}{0.400pt}}
\put(1003.0,585.0){\rule[-0.200pt]{4.818pt}{0.400pt}}
\put(843.0,544.0){\rule[-0.200pt]{0.400pt}{7.709pt}}
\put(833.0,544.0){\rule[-0.200pt]{4.818pt}{0.400pt}}
\put(833.0,576.0){\rule[-0.200pt]{4.818pt}{0.400pt}}
\put(729.0,527.0){\rule[-0.200pt]{0.400pt}{6.263pt}}
\put(719.0,527.0){\rule[-0.200pt]{4.818pt}{0.400pt}}
\put(719.0,553.0){\rule[-0.200pt]{4.818pt}{0.400pt}}
\put(1183.0,569.0){\rule[-0.200pt]{0.400pt}{5.782pt}}
\put(1173.0,569.0){\rule[-0.200pt]{4.818pt}{0.400pt}}
\put(1013,577){\makebox(0,0){$\times$}}
\put(843,560){\makebox(0,0){$\times$}}
\put(729,540){\makebox(0,0){$\times$}}
\put(1183,581){\makebox(0,0){$\times$}}
\put(1173.0,593.0){\rule[-0.200pt]{4.818pt}{0.400pt}}
\put(1183.0,721.0){\rule[-0.200pt]{0.400pt}{4.095pt}}
\put(1173.0,721.0){\rule[-0.200pt]{4.818pt}{0.400pt}}
\put(1173.0,738.0){\rule[-0.200pt]{4.818pt}{0.400pt}}
\put(1013.0,698.0){\rule[-0.200pt]{0.400pt}{3.613pt}}
\put(1003.0,698.0){\rule[-0.200pt]{4.818pt}{0.400pt}}
\put(1003.0,713.0){\rule[-0.200pt]{4.818pt}{0.400pt}}
\put(843.0,679.0){\rule[-0.200pt]{0.400pt}{8.191pt}}
\put(833.0,679.0){\rule[-0.200pt]{4.818pt}{0.400pt}}
\put(833.0,713.0){\rule[-0.200pt]{4.818pt}{0.400pt}}
\put(729.0,656.0){\rule[-0.200pt]{0.400pt}{8.191pt}}
\put(719.0,656.0){\rule[-0.200pt]{4.818pt}{0.400pt}}
\put(1183,730){\makebox(0,0){$\triangle$}}
\put(1013,705){\makebox(0,0){$\triangle$}}
\put(843,696){\makebox(0,0){$\triangle$}}
\put(729,673){\makebox(0,0){$\triangle$}}
\put(719.0,690.0){\rule[-0.200pt]{4.818pt}{0.400pt}}
\end{picture}
\caption{Inverse integral autocorrelation time in 
units of $g^2T$ plotted against
$g^2Ta$ for
$\langle {\bf D}\times{\bf B}({\bf x},t)\cdot{\bf D}\times{\bf B}({\bf x},0)
\rangle$ (triangles), $\langle {\bf B}({\bf x},t\cdot{\bf B}({\bf x},0)\rangle$
(crosses), and
$\langle B_i^2({\bf x},t)B_i^2({\bf x},0)\rangle-
\langle B_i^2({\bf x},t)\rangle\langle B_i^2({\bf x},0)\rangle$ (squares).}
\label{tauinv}
\end{figure}
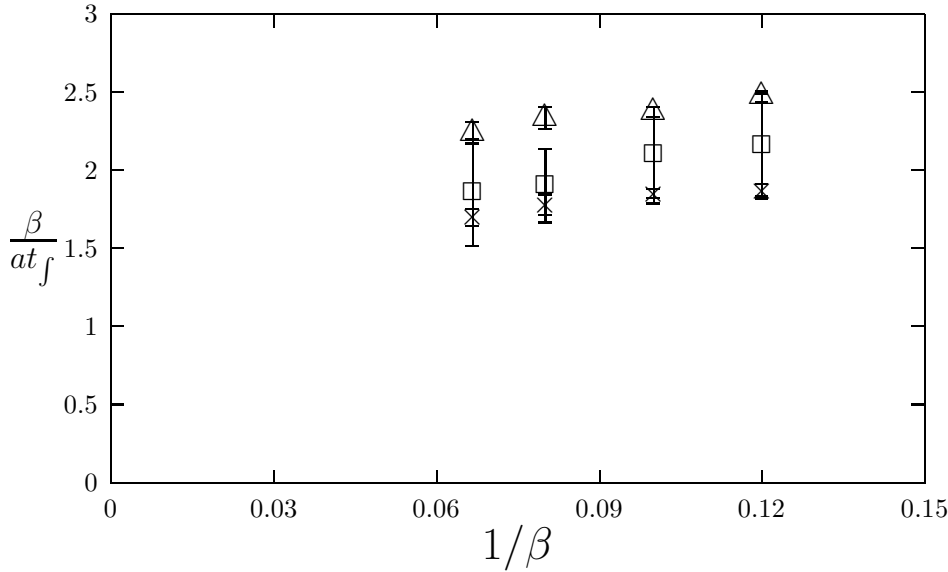

To further quantify this behavior of the time scales, we introduce the integral
autocorrelation time defined for an autocorrelator $C(t)$ as
$$t_{\int}\equiv (C(0))^{-1}\left(\int_0^\infty C(t){\rm d}t\right),$$
where in our numerical estimates the upper limit of 
integration is replaced by a
finite value $t_u$, for which $C(t_u)/C(0)\ll 1$. In Figure \ref{tauinv} we 
plot the
dimensionless quantity $4/(g^2Tt_{\int})$ as a function of $1/\beta=g^2Ta/4$. 
Remarkably, in all three cases $t_{\int}$ turns out to be of the order of 
$g^2T$ and shows little dependence on the lattice spacing throughout the range
considered. There is therefore no evidence that in this range of the lattice 
spacings our cooled autocorrelators follow the ASY -- B\"odeker scenario,
wherein the expected behavior is 
$t_{\int}\propto 1/(g^4T^2a)$, up to logarithmic
corrections.

The autocorrelator $\langle B_i^2({\bf x},t)B_i^2({\bf x},0)\rangle$ turned out
to be a much noisier quantity than the other two, 
resulting in much larger error
bars despite comparable sample sizes in all the three cases. Apart from this 
difference, the integral autocorrelation times behave very similarly for the
gauge-covariant and for the gauge-invariant quantities, even though the 
gauge-invariant autocorrelator of the former necessarily involves a straight
adjoint Wilson line connecting the points ${\bf x},t$ and ${\bf x},0$.

\begin{figure}[hbt]
%\input pl
% GNUPLOT: LaTeX picture
\setlength{\unitlength}{0.240900pt}
\ifx\plotpoint\undefined\newsavebox{\plotpoint}\fi
\sbox{\plotpoint}{\rule[-0.200pt]{0.400pt}{0.400pt}}%
\begin{picture}(1500,900)(0,0)
\font\gnuplot=cmr10 at 10pt
\gnuplot
\sbox{\plotpoint}{\rule[-0.200pt]{0.400pt}{0.400pt}}%
\put(181.0,123.0){\rule[-0.200pt]{4.818pt}{0.400pt}}
\put(161,123){\makebox(0,0)[r]{-0.2}}
\put(1419.0,123.0){\rule[-0.200pt]{4.818pt}{0.400pt}}
\put(181.0,228.0){\rule[-0.200pt]{4.818pt}{0.400pt}}
\put(161,228){\makebox(0,0)[r]{0}}
\put(1419.0,228.0){\rule[-0.200pt]{4.818pt}{0.400pt}}
\put(181.0,334.0){\rule[-0.200pt]{4.818pt}{0.400pt}}
\put(161,334){\makebox(0,0)[r]{0.2}}
\put(1419.0,334.0){\rule[-0.200pt]{4.818pt}{0.400pt}}
\put(181.0,439.0){\rule[-0.200pt]{4.818pt}{0.400pt}}
\put(161,439){\makebox(0,0)[r]{0.4}}
\put(1419.0,439.0){\rule[-0.200pt]{4.818pt}{0.400pt}}
\put(181.0,544.0){\rule[-0.200pt]{4.818pt}{0.400pt}}
\put(161,544){\makebox(0,0)[r]{0.6}}
\put(1419.0,544.0){\rule[-0.200pt]{4.818pt}{0.400pt}}
\put(181.0,649.0){\rule[-0.200pt]{4.818pt}{0.400pt}}
\put(161,649){\makebox(0,0)[r]{0.8}}
\put(1419.0,649.0){\rule[-0.200pt]{4.818pt}{0.400pt}}
\put(181.0,755.0){\rule[-0.200pt]{4.818pt}{0.400pt}}
\put(161,755){\makebox(0,0)[r]{1}}
\put(1419.0,755.0){\rule[-0.200pt]{4.818pt}{0.400pt}}
\put(181.0,860.0){\rule[-0.200pt]{4.818pt}{0.400pt}}
\put(161,860){\makebox(0,0)[r]{1.2}}
\put(1419.0,860.0){\rule[-0.200pt]{4.818pt}{0.400pt}}
\put(181.0,123.0){\rule[-0.200pt]{0.400pt}{4.818pt}}
\put(181,82){\makebox(0,0){0}}
\put(181.0,840.0){\rule[-0.200pt]{0.400pt}{4.818pt}}
\put(391.0,123.0){\rule[-0.200pt]{0.400pt}{4.818pt}}
\put(391,82){\makebox(0,0){1}}
\put(391.0,840.0){\rule[-0.200pt]{0.400pt}{4.818pt}}
\put(600.0,123.0){\rule[-0.200pt]{0.400pt}{4.818pt}}
\put(600,82){\makebox(0,0){2}}
\put(600.0,840.0){\rule[-0.200pt]{0.400pt}{4.818pt}}
\put(810.0,123.0){\rule[-0.200pt]{0.400pt}{4.818pt}}
\put(810,82){\makebox(0,0){3}}
\put(810.0,840.0){\rule[-0.200pt]{0.400pt}{4.818pt}}
\put(1020.0,123.0){\rule[-0.200pt]{0.400pt}{4.818pt}}
\put(1020,82){\makebox(0,0){4}}
\put(1020.0,840.0){\rule[-0.200pt]{0.400pt}{4.818pt}}
\put(1229.0,123.0){\rule[-0.200pt]{0.400pt}{4.818pt}}
\put(1229,82){\makebox(0,0){5}}
\put(1229.0,840.0){\rule[-0.200pt]{0.400pt}{4.818pt}}
\put(181.0,123.0){\rule[-0.200pt]{303.052pt}{0.400pt}}
\put(1439.0,123.0){\rule[-0.200pt]{0.400pt}{177.543pt}}
\put(181.0,860.0){\rule[-0.200pt]{303.052pt}{0.400pt}}
\put(40,491){\makebox(0,0){\Large{${C(t)}\over{C(0)}$}}}
\put(810,21){\makebox(0,0){\Large{$ta$}}}
\put(181.0,123.0){\rule[-0.200pt]{0.400pt}{177.543pt}}
\put(181.0,754.0){\usebox{\plotpoint}}
\put(171.0,754.0){\rule[-0.200pt]{4.818pt}{0.400pt}}
\put(171.0,755.0){\rule[-0.200pt]{4.818pt}{0.400pt}}
\put(307,655){\usebox{\plotpoint}}
\put(297.0,655.0){\rule[-0.200pt]{4.818pt}{0.400pt}}
\put(297.0,655.0){\rule[-0.200pt]{4.818pt}{0.400pt}}
\put(433,505){\usebox{\plotpoint}}
\put(423.0,505.0){\rule[-0.200pt]{4.818pt}{0.400pt}}
\put(423.0,505.0){\rule[-0.200pt]{4.818pt}{0.400pt}}
\put(558,396){\usebox{\plotpoint}}
\put(548.0,396.0){\rule[-0.200pt]{4.818pt}{0.400pt}}
\put(548.0,396.0){\rule[-0.200pt]{4.818pt}{0.400pt}}
\put(684.0,317.0){\usebox{\plotpoint}}
\put(674.0,317.0){\rule[-0.200pt]{4.818pt}{0.400pt}}
\put(674.0,318.0){\rule[-0.200pt]{4.818pt}{0.400pt}}
\put(810,267){\usebox{\plotpoint}}
\put(800.0,267.0){\rule[-0.200pt]{4.818pt}{0.400pt}}
\put(800.0,267.0){\rule[-0.200pt]{4.818pt}{0.400pt}}
\put(936,240){\usebox{\plotpoint}}
\put(926.0,240.0){\rule[-0.200pt]{4.818pt}{0.400pt}}
\put(926.0,240.0){\rule[-0.200pt]{4.818pt}{0.400pt}}
\put(1062,228){\usebox{\plotpoint}}
\put(1052.0,228.0){\rule[-0.200pt]{4.818pt}{0.400pt}}
\put(1052.0,228.0){\rule[-0.200pt]{4.818pt}{0.400pt}}
\put(1187,223){\usebox{\plotpoint}}
\put(1177.0,223.0){\rule[-0.200pt]{4.818pt}{0.400pt}}
\put(1177.0,223.0){\rule[-0.200pt]{4.818pt}{0.400pt}}
\put(1313,222){\usebox{\plotpoint}}
\put(1303.0,222.0){\rule[-0.200pt]{4.818pt}{0.400pt}}
\put(181,755){\raisebox{-.8pt}{\makebox(0,0){$\Box$}}}
\put(307,655){\raisebox{-.8pt}{\makebox(0,0){$\Box$}}}
\put(433,505){\raisebox{-.8pt}{\makebox(0,0){$\Box$}}}
\put(558,396){\raisebox{-.8pt}{\makebox(0,0){$\Box$}}}
\put(684,318){\raisebox{-.8pt}{\makebox(0,0){$\Box$}}}
\put(810,267){\raisebox{-.8pt}{\makebox(0,0){$\Box$}}}
\put(936,240){\raisebox{-.8pt}{\makebox(0,0){$\Box$}}}
\put(1062,228){\raisebox{-.8pt}{\makebox(0,0){$\Box$}}}
\put(1187,223){\raisebox{-.8pt}{\makebox(0,0){$\Box$}}}
\put(1313,222){\raisebox{-.8pt}{\makebox(0,0){$\Box$}}}
\put(1303.0,222.0){\rule[-0.200pt]{4.818pt}{0.400pt}}
\put(181.0,754.0){\usebox{\plotpoint}}
\put(171.0,754.0){\rule[-0.200pt]{4.818pt}{0.400pt}}
\put(171.0,755.0){\rule[-0.200pt]{4.818pt}{0.400pt}}
\put(307,662){\usebox{\plotpoint}}
\put(297.0,662.0){\rule[-0.200pt]{4.818pt}{0.400pt}}
\put(297.0,662.0){\rule[-0.200pt]{4.818pt}{0.400pt}}
\put(433,517){\usebox{\plotpoint}}
\put(423.0,517.0){\rule[-0.200pt]{4.818pt}{0.400pt}}
\put(423.0,517.0){\rule[-0.200pt]{4.818pt}{0.400pt}}
\put(558,407){\usebox{\plotpoint}}
\put(548.0,407.0){\rule[-0.200pt]{4.818pt}{0.400pt}}
\put(548.0,407.0){\rule[-0.200pt]{4.818pt}{0.400pt}}
\put(684,324){\usebox{\plotpoint}}
\put(674.0,324.0){\rule[-0.200pt]{4.818pt}{0.400pt}}
\put(674.0,324.0){\rule[-0.200pt]{4.818pt}{0.400pt}}
\put(810,270){\usebox{\plotpoint}}
\put(800.0,270.0){\rule[-0.200pt]{4.818pt}{0.400pt}}
\put(800.0,270.0){\rule[-0.200pt]{4.818pt}{0.400pt}}
\put(936,242){\usebox{\plotpoint}}
\put(926.0,242.0){\rule[-0.200pt]{4.818pt}{0.400pt}}
\put(926.0,242.0){\rule[-0.200pt]{4.818pt}{0.400pt}}
\put(1062,229){\usebox{\plotpoint}}
\put(1052.0,229.0){\rule[-0.200pt]{4.818pt}{0.400pt}}
\put(1052.0,229.0){\rule[-0.200pt]{4.818pt}{0.400pt}}
\put(1187,223){\usebox{\plotpoint}}
\put(1177.0,223.0){\rule[-0.200pt]{4.818pt}{0.400pt}}
\put(1177.0,223.0){\rule[-0.200pt]{4.818pt}{0.400pt}}
\put(1313,222){\usebox{\plotpoint}}
\put(1303.0,222.0){\rule[-0.200pt]{4.818pt}{0.400pt}}
\put(181,755){\makebox(0,0){$\times$}}
\put(307,662){\makebox(0,0){$\times$}}
\put(433,517){\makebox(0,0){$\times$}}
\put(558,407){\makebox(0,0){$\times$}}
\put(684,324){\makebox(0,0){$\times$}}
\put(810,270){\makebox(0,0){$\times$}}
\put(936,242){\makebox(0,0){$\times$}}
\put(1062,229){\makebox(0,0){$\times$}}
\put(1187,223){\makebox(0,0){$\times$}}
\put(1313,222){\makebox(0,0){$\times$}}
\put(1303.0,222.0){\rule[-0.200pt]{4.818pt}{0.400pt}}
\put(181,755){\usebox{\plotpoint}}
\put(171.0,755.0){\rule[-0.200pt]{4.818pt}{0.400pt}}
\put(171.0,755.0){\rule[-0.200pt]{4.818pt}{0.400pt}}
\put(307,668){\usebox{\plotpoint}}
\put(297.0,668.0){\rule[-0.200pt]{4.818pt}{0.400pt}}
\put(297.0,668.0){\rule[-0.200pt]{4.818pt}{0.400pt}}
\put(433,530){\usebox{\plotpoint}}
\put(423.0,530.0){\rule[-0.200pt]{4.818pt}{0.400pt}}
\put(423.0,530.0){\rule[-0.200pt]{4.818pt}{0.400pt}}
\put(558,417){\usebox{\plotpoint}}
\put(548.0,417.0){\rule[-0.200pt]{4.818pt}{0.400pt}}
\put(548.0,417.0){\rule[-0.200pt]{4.818pt}{0.400pt}}
\put(684,330){\usebox{\plotpoint}}
\put(674.0,330.0){\rule[-0.200pt]{4.818pt}{0.400pt}}
\put(674.0,330.0){\rule[-0.200pt]{4.818pt}{0.400pt}}
\put(810,274){\usebox{\plotpoint}}
\put(800.0,274.0){\rule[-0.200pt]{4.818pt}{0.400pt}}
\put(800.0,274.0){\rule[-0.200pt]{4.818pt}{0.400pt}}
\put(936,245){\usebox{\plotpoint}}
\put(926.0,245.0){\rule[-0.200pt]{4.818pt}{0.400pt}}
\put(926.0,245.0){\rule[-0.200pt]{4.818pt}{0.400pt}}
\put(1062,231){\usebox{\plotpoint}}
\put(1052.0,231.0){\rule[-0.200pt]{4.818pt}{0.400pt}}
\put(1052.0,231.0){\rule[-0.200pt]{4.818pt}{0.400pt}}
\put(1187.0,224.0){\usebox{\plotpoint}}
\put(1177.0,224.0){\rule[-0.200pt]{4.818pt}{0.400pt}}
\put(1177.0,225.0){\rule[-0.200pt]{4.818pt}{0.400pt}}
\put(1313,223){\usebox{\plotpoint}}
\put(1303.0,223.0){\rule[-0.200pt]{4.818pt}{0.400pt}}
\put(181,755){\makebox(0,0){$\triangle$}}
\put(307,668){\makebox(0,0){$\triangle$}}
\put(433,530){\makebox(0,0){$\triangle$}}
\put(558,417){\makebox(0,0){$\triangle$}}
\put(684,330){\makebox(0,0){$\triangle$}}
\put(810,274){\makebox(0,0){$\triangle$}}
\put(936,245){\makebox(0,0){$\triangle$}}
\put(1062,231){\makebox(0,0){$\triangle$}}
\put(1187,224){\makebox(0,0){$\triangle$}}
\put(1313,223){\makebox(0,0){$\triangle$}}
\put(1303.0,223.0){\rule[-0.200pt]{4.818pt}{0.400pt}}
\end{picture}
\caption{The autocorrelator
$\langle {\bf D}\cdot{\bf E}({\bf x},t){\bf D}\cdot{\bf E}({\bf x},0)\rangle$
versus time
$t$ in  lattice units for $\beta=10$ (crosses), 
$\beta=12.0$ (squares), and $\beta=14.0$ (triangles) Note the complete or 
partial overlap of the data points.}
\label{pl}
\end{figure}
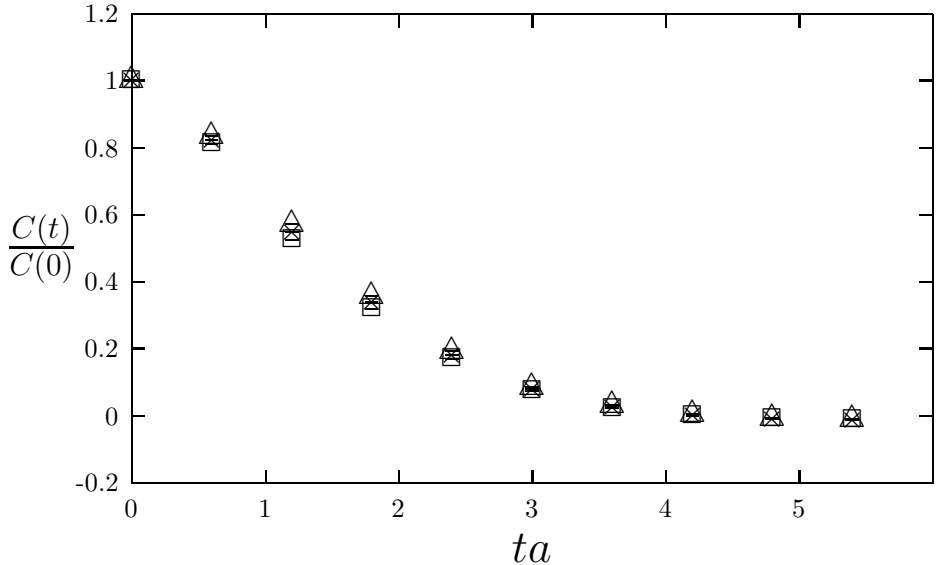

Next, we consider the color charge autocorrelator
$\langle {\bf D}\cdot{\bf E}({\bf x},t){\bf D}\cdot{\bf E}({\bf x},0)\rangle$.
As shown in Figure \ref{pl}, this quantity is strikingly different from the 
magnetic-field autocorrelators considered earlier. 
The time scale for the color 
charge correlation is proportional to the lattice spacing and does not depend
on $g^2T$. 

This result can be contrasted with perturbative predictions.
One would expect that the color-charge autocorrelator is dominated by
the plasmon mode, whose frequency in the classical theory is of the 
order $g\sqrt{T/a}$ and whose decay rate is of the order $g^2T$. We observe
none of these properties in the range of lattice spacings considered.

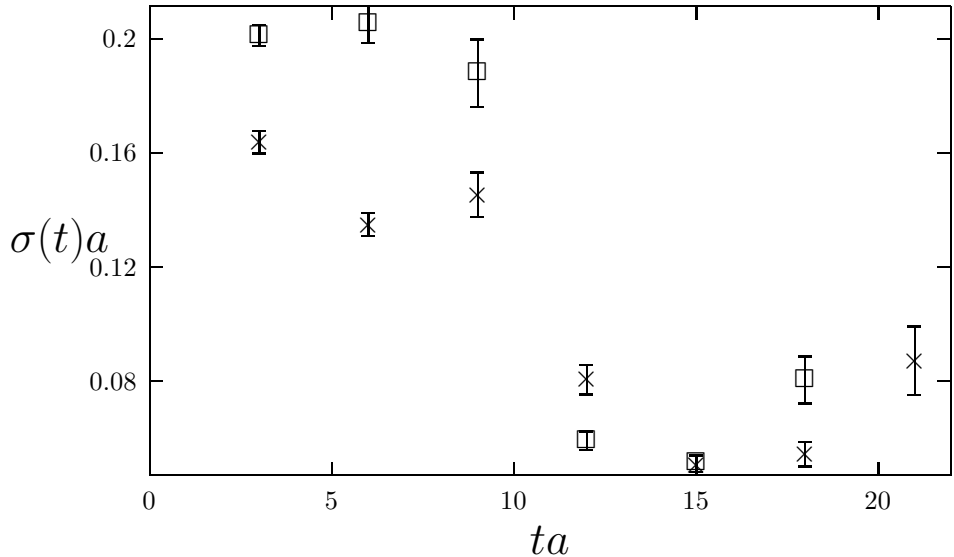
\begin{figure}[hbt]
%\input cc
% GNUPLOT: LaTeX picture
\setlength{\unitlength}{0.240900pt}
\ifx\plotpoint\undefined\newsavebox{\plotpoint}\fi
\sbox{\plotpoint}{\rule[-0.200pt]{0.400pt}{0.400pt}}%
\begin{picture}(1500,900)(0,0)
\font\gnuplot=cmr10 at 10pt
\gnuplot
\sbox{\plotpoint}{\rule[-0.200pt]{0.400pt}{0.400pt}}%
\put(181.0,270.0){\rule[-0.200pt]{4.818pt}{0.400pt}}
\put(161,270){\makebox(0,0)[r]{0.08}}
\put(1419.0,270.0){\rule[-0.200pt]{4.818pt}{0.400pt}}
\put(181.0,449.0){\rule[-0.200pt]{4.818pt}{0.400pt}}
\put(161,449){\makebox(0,0)[r]{0.12}}
\put(1419.0,449.0){\rule[-0.200pt]{4.818pt}{0.400pt}}
\put(181.0,629.0){\rule[-0.200pt]{4.818pt}{0.400pt}}
\put(161,629){\makebox(0,0)[r]{0.16}}
\put(1419.0,629.0){\rule[-0.200pt]{4.818pt}{0.400pt}}
\put(181.0,808.0){\rule[-0.200pt]{4.818pt}{0.400pt}}
\put(161,808){\makebox(0,0)[r]{0.2}}
\put(1419.0,808.0){\rule[-0.200pt]{4.818pt}{0.400pt}}
\put(181.0,123.0){\rule[-0.200pt]{0.400pt}{4.818pt}}
\put(181,82){\makebox(0,0){0}}
\put(181.0,840.0){\rule[-0.200pt]{0.400pt}{4.818pt}}
\put(467.0,123.0){\rule[-0.200pt]{0.400pt}{4.818pt}}
\put(467,82){\makebox(0,0){5}}
\put(467.0,840.0){\rule[-0.200pt]{0.400pt}{4.818pt}}
\put(753.0,123.0){\rule[-0.200pt]{0.400pt}{4.818pt}}
\put(753,82){\makebox(0,0){10}}
\put(753.0,840.0){\rule[-0.200pt]{0.400pt}{4.818pt}}
\put(1039.0,123.0){\rule[-0.200pt]{0.400pt}{4.818pt}}
\put(1039,82){\makebox(0,0){15}}
\put(1039.0,840.0){\rule[-0.200pt]{0.400pt}{4.818pt}}
\put(1325.0,123.0){\rule[-0.200pt]{0.400pt}{4.818pt}}
\put(1325,82){\makebox(0,0){20}}
\put(1325.0,840.0){\rule[-0.200pt]{0.400pt}{4.818pt}}
\put(181.0,123.0){\rule[-0.200pt]{303.052pt}{0.400pt}}
\put(1439.0,123.0){\rule[-0.200pt]{0.400pt}{177.543pt}}
\put(181.0,860.0){\rule[-0.200pt]{303.052pt}{0.400pt}}
\put(40,491){\makebox(0,0){\Large{$\sigma(t)a$}}}
\put(810,21){\makebox(0,0){\Large{$ta$}}}
\put(181.0,123.0){\rule[-0.200pt]{0.400pt}{177.543pt}}
\put(353.0,797.0){\rule[-0.200pt]{0.400pt}{7.950pt}}
\put(343.0,797.0){\rule[-0.200pt]{4.818pt}{0.400pt}}
\put(343.0,830.0){\rule[-0.200pt]{4.818pt}{0.400pt}}
\put(524.0,802.0){\rule[-0.200pt]{0.400pt}{13.972pt}}
\put(514.0,802.0){\rule[-0.200pt]{4.818pt}{0.400pt}}
\put(514.0,860.0){\rule[-0.200pt]{4.818pt}{0.400pt}}
\put(696.0,701.0){\rule[-0.200pt]{0.400pt}{25.535pt}}
\put(686.0,701.0){\rule[-0.200pt]{4.818pt}{0.400pt}}
\put(686.0,807.0){\rule[-0.200pt]{4.818pt}{0.400pt}}
\put(867.0,162.0){\rule[-0.200pt]{0.400pt}{6.986pt}}
\put(857.0,162.0){\rule[-0.200pt]{4.818pt}{0.400pt}}
\put(857.0,191.0){\rule[-0.200pt]{4.818pt}{0.400pt}}
\put(1039.0,127.0){\rule[-0.200pt]{0.400pt}{6.504pt}}
\put(1029.0,127.0){\rule[-0.200pt]{4.818pt}{0.400pt}}
\put(1029.0,154.0){\rule[-0.200pt]{4.818pt}{0.400pt}}
\put(1210.0,235.0){\rule[-0.200pt]{0.400pt}{17.827pt}}
\put(1200.0,235.0){\rule[-0.200pt]{4.818pt}{0.400pt}}
\put(353,813){\raisebox{-.8pt}{\makebox(0,0){$\Box$}}}
\put(524,831){\raisebox{-.8pt}{\makebox(0,0){$\Box$}}}
\put(696,754){\raisebox{-.8pt}{\makebox(0,0){$\Box$}}}
\put(867,176){\raisebox{-.8pt}{\makebox(0,0){$\Box$}}}
\put(1039,141){\raisebox{-.8pt}{\makebox(0,0){$\Box$}}}
\put(1210,272){\raisebox{-.8pt}{\makebox(0,0){$\Box$}}}
\put(1200.0,309.0){\rule[-0.200pt]{4.818pt}{0.400pt}}
\put(353.0,628.0){\rule[-0.200pt]{0.400pt}{8.431pt}}
\put(343.0,628.0){\rule[-0.200pt]{4.818pt}{0.400pt}}
\put(343.0,663.0){\rule[-0.200pt]{4.818pt}{0.400pt}}
\put(524.0,498.0){\rule[-0.200pt]{0.400pt}{8.672pt}}
\put(514.0,498.0){\rule[-0.200pt]{4.818pt}{0.400pt}}
\put(514.0,534.0){\rule[-0.200pt]{4.818pt}{0.400pt}}
\put(696.0,528.0){\rule[-0.200pt]{0.400pt}{16.863pt}}
\put(686.0,528.0){\rule[-0.200pt]{4.818pt}{0.400pt}}
\put(686.0,598.0){\rule[-0.200pt]{4.818pt}{0.400pt}}
\put(867.0,249.0){\rule[-0.200pt]{0.400pt}{11.322pt}}
\put(857.0,249.0){\rule[-0.200pt]{4.818pt}{0.400pt}}
\put(857.0,296.0){\rule[-0.200pt]{4.818pt}{0.400pt}}
\put(1039.0,123.0){\rule[-0.200pt]{0.400pt}{7.227pt}}
\put(1029.0,123.0){\rule[-0.200pt]{4.818pt}{0.400pt}}
\put(1029.0,153.0){\rule[-0.200pt]{4.818pt}{0.400pt}}
\put(1210.0,136.0){\rule[-0.200pt]{0.400pt}{9.154pt}}
\put(1200.0,136.0){\rule[-0.200pt]{4.818pt}{0.400pt}}
\put(1200.0,174.0){\rule[-0.200pt]{4.818pt}{0.400pt}}
\put(1382.0,248.0){\rule[-0.200pt]{0.400pt}{26.017pt}}
\put(1372.0,248.0){\rule[-0.200pt]{4.818pt}{0.400pt}}
\put(353,646){\makebox(0,0){$\times$}}
\put(524,516){\makebox(0,0){$\times$}}
\put(696,563){\makebox(0,0){$\times$}}
\put(867,273){\makebox(0,0){$\times$}}
\put(1039,138){\makebox(0,0){$\times$}}
\put(1210,155){\makebox(0,0){$\times$}}
\put(1382,302){\makebox(0,0){$\times$}}
\put(1372.0,356.0){\rule[-0.200pt]{4.818pt}{0.400pt}}
\end{picture}
\caption{The color conductivity in lattice units defined as in Eq. \ref{2.a3}
versus time
$t$ in lattice units for $\beta=10$ (crosses) and
$\beta=12.0$ (squares).}
\label{cc}
\end{figure}

Finally, we attempted to determine color conductivity $\sigma$, as defined in 
the Introduction, using Eq. \ref{2.a3}. As Figure \ref{cc} demonstrates, 
this attempt
failed in two ways. First of all, $\sigma(t)$ does not appear to 
approach a constant for times in excess of the expected autocorrelation time
of the noise term ${\bf \xi}$ 
(and far in excess of the measured autocorrelation
time of the noise). Secondly, the numerical value of $\sigma(t)$ is very
small (less than $0.25/a$) compared to the value expected in the ASY scenario 
($a\sigma\approx 15$). Given this small value of $\sigma(t)$, it is not clear 
how neglecting the $\dot{\bf E}$ term in Eq. \ref{1.1} can be justified.

\section{Conclusion}

In summary, we see no sign of the ASY-B\"odeker 
scenario in the classical SU(2) theory, in the regime roughly
corresponding to the electroweak scale. This negative finding does not rule
out the ASY-B\"odeker scenario in general, but one is led to question its
applicability outside the asymptotically weak-coupling regime, which in the
classical lattice theory corresponds to asymptotically small lattice spacing.

Neither our findings contradict earlier numerical data for the classical 
Yang-Mills theory, in particular, the sphaleron rate measurement by
Moore and Rummukainen \cite{quant2}. Their simulation was performed in 
the range of 
couplings (or lattice spacings) which overlaps the one considered here.
Results of that work are consistent with the zero continuum limit of the rate,
as predicted by ASY and by B\"odeker. But they do not rule out a finite
classical rate in the continuum. 

This brings us to the following methodological remark. The sphaleron transition
rate is a very important quantity in its own right, worthy of a careful 
numerical study. 
However, such study may not be the optimal way to test the theory of ASY and
B\"odeker. Precisely because the sphaleron rate is an essentially 
nonperturbative quantity, there is no easy way to disentangle the 
perturbative ASY-B\"odeker damping from genuinely nonperturbative effects.
To compound the difficulty, topology on a lattice is ill-defined and requires
special treatment. By contrast, a study like the one reported here attempts
to make a direct contact with perturbation theory, by measuring objects
such as color conductivity. It therefore may be better suited for testing
perturbative predictions.

\vspace{24pt}

\noindent
{\large \bf  Acknowledgement.}
The authors are grateful to P. Arnold, D. B\"odeker, and L. Yaffe for 
illuminating discussions.
The work of J.A. and K.N.A. was supported in part by
``MaPhySto'', {\it Center of Mathematical Physics and Stochastics}--
financed by the Danish National Research Foundation.
J.A. and A.K. acknowledge the support of the Portuguese Funda\c c\~ao para a
Ci\^encia e a Tecnologia, under the research grants
CERN/P/FIS/1203/98 and CERN/P/FIS/15196/1999. 
K.N.A.'s research was partially supported by the RTN grants
HPRN-CT-2000-00122 and HPRN-CT-2000-00131 and a National Fellowship
Foundation of Greece (IKY) postdoctoral fellowship.
Finally, the authors 
acknowledge the support of the Danish Natural Science Council in 
making the simulations at the UNI-C computer center possible.


\begin{thebibliography}{9}

\bibitem{krs}
V.A. Kuzmin, V.A. Rubakov and M.E. Shaposhnikov,
Phys. Lett. {\bf B155} (1985) 36.
\bibitem{grs}
Yu. Grigoriev, V.A. Rubakov and M.E. Shaposhnikov,
Phys. Lett. {\bf B216} (1989) 172. 

\bibitem{aaps} J. Ambj\o rn, T. Askgaard, H. Porter  and M.E. Shaposhnikov,
Phys. Lett. {\bf B244} (1990) 479; Nucl.Phys. {\bf B353} (1991) 346.

\bibitem{su2rate}J. Ambj{\o}rn and A. Krasnitz, Phys. Lett. {\bf B362} (1995)
97, hep-ph/9508202.
\bibitem{thalgs}A. Krasnitz, Nucl. Phys. {\bf B455} (1995) 320, hep-lat/9507025.

\bibitem{quant1} D. B\"odeker, Guy D. Moore and K. Rummukainen,
Phys.\ Rev.\ {\bf D61} (2000) 056003, hep-ph/9907545.
\bibitem{quant2} Guy D.\ Moore and Kari Rummukainen,
Phys.\ Rev.\ {\bf D61} (2000) 105008, hep-ph/9906259.
\bibitem{quant3} Guy D. Moore, Nucl.\ Phys\. {\bf B568} (2000) 367,
hep-ph/9810313;
Phys.\ Lett.\ {\bf B412} (1997) 359, hep-ph/9705248.
\bibitem{quant4} J.\ Ambj{\o}rn, A.\ Krasnitz, Nucl.\ Phys.\ {\bf B506} (1997)
387, hep-ph/9705380.
\bibitem{quant5} Guy D. Moore, Chao-Ran Hu and Berndt M\"uller,
Phys.\ Rev.\ {\bf D58} (1998) 045001, hep-ph/9710436. 
\bibitem{quant6} Guy D. Moore and Neil Turok, Phys.\ Rev.\ {\bf D56} (1997) 
6533, hep-ph/9703266; Phys.\ Rev.\ {\bf D55} (1997) 6538, hep-ph/9608350.
\bibitem{quant7} M. Salle, J. Smit, J.C. Vink, {\it Scalar field dynamics: 
classical, quantum and in between}, hep-ph/0009120;\\
Gert Aarts and Jan Smit, Nucl.\ Phys.\ {\bf B511} (1998) 451, hep-ph/9707342;\\
Wai-Hung Tang and  Jan Smit,  Nucl.\ Phys.\ {\bf B510} (1998) 401, 
hep-lat/9702017; Nucl.\ Phys.\ {\bf B482} (1996) 265, hep-lat/9605016. 

\bibitem{topology}J. Ambj\o rn and K. Farakos, Phys. Lett. 
{\bf B294} (1992) 248, hep-lat/9207020;\\
J. Ambj\o rn, K. Farakos, S. Hands, 
G. Koutsoumbas and G. Thorleifsson, Nucl. Phys. {\bf B425} (1994) 39, 
hep-lat/9401028.

\bibitem{bodeker}
D.\ B\"odeker, Phys.\ Lett.\ {\bf B426} (1998) 351,  hep-ph/9801430;  
Nucl.\ Phys.\ {\bf B566} (2000) 402, hep-ph/9903478;  
Nucl.\ Phys.\ {\bf B559} (1999) 502, hep-ph/9905239; 
{\it A local Langevin equation for slow 
long-distance modes of hot non-abelian gauge fields},
NBI-HE-00-46, hep-ph/0012304.


\bibitem{asy} P.\ Arnold, D.\ Son, L.\ G.\ Yaffe, Phys.\ Rev.\ 
{\bf D55} (1997) 6264, hep-ph/9609481; 
Phys.\ Rev.\ {\bf D57} (1998) 1178, hep-ph/9709449; 
Phys.\ Rev.\ {\bf D59} (1999) 105020, hep-ph/9810216.
\bibitem{as}P.\ Arnold, L.\ G.\ Yaffe, Phys. Rev. {\bf D62} (2000) 125014,
hep-ph/9912306; {\it ibid} 125013, hep-ph/9912305.

\bibitem{aals}J.\ Ambj{\o}rn, M.L.\ Laursen and M.E. Shaposhnikov,
Nucl.\ Phys.\ {\bf B316} (1989) 483; Phys.\ Lett.\ {\bf B197} (1987) 49. 

\bibitem{arnold1}P.\ Arnold, Phys.\ Rev.\ {\bf E61} (2000) 6099, 
hep-ph/9912209.  



\end{thebibliography}
\end{document}